\documentclass[11pt]{article}
\usepackage{amsmath,amsfonts,amsthm,amsbsy} 
 \usepackage{graphicx}
\DeclareGraphicsExtensions{.eps}

\numberwithin{equation}{section}

\newcommand\figPath{./figures/}

 \newcommand{\asep}{\includegraphics{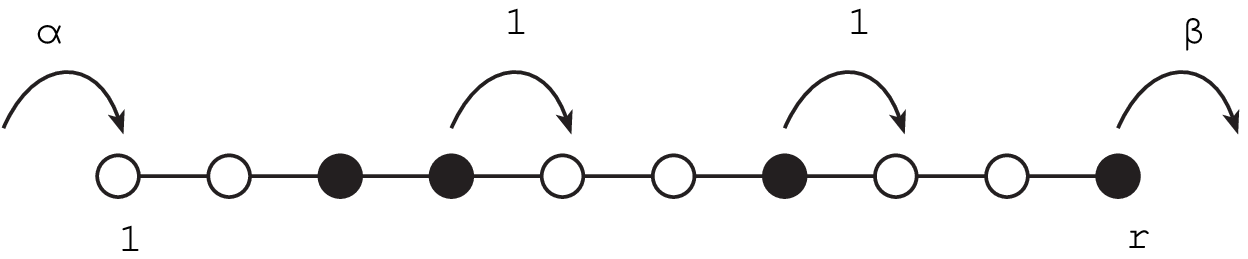}}
\newcommand{\ballot}{\includegraphics{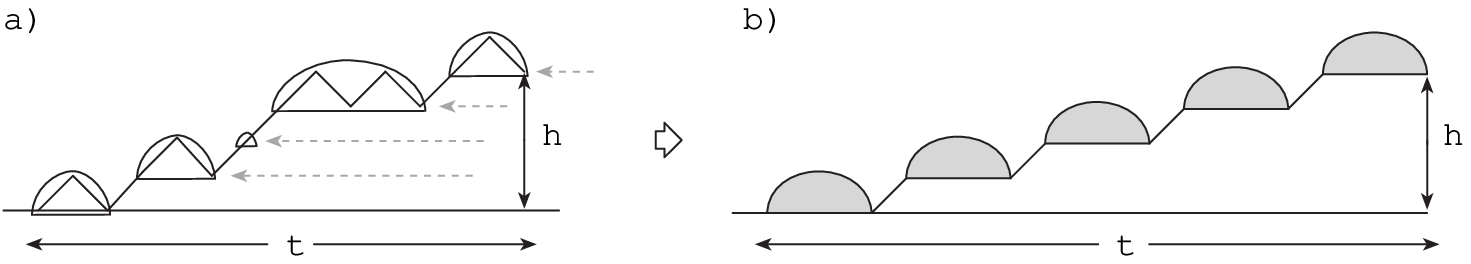}}
\newcommand{\returns}{\includegraphics{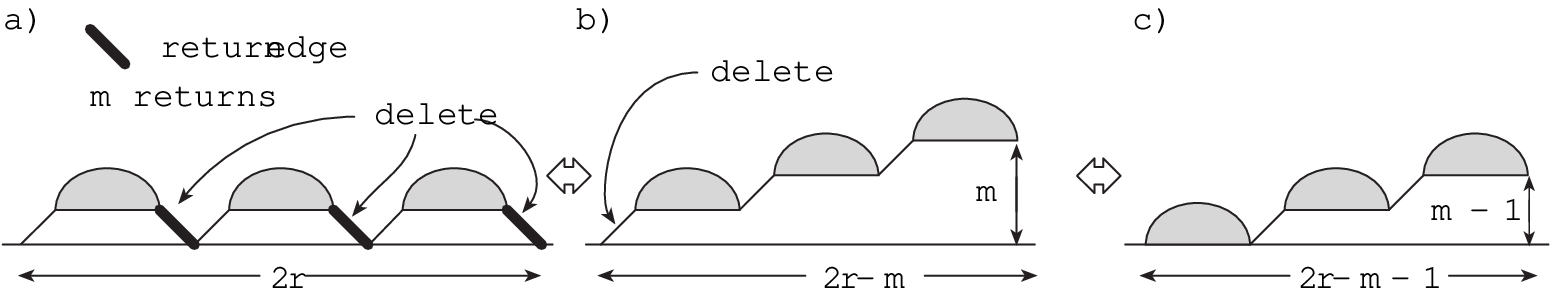}}
\newcommand{\markedContact}{\includegraphics{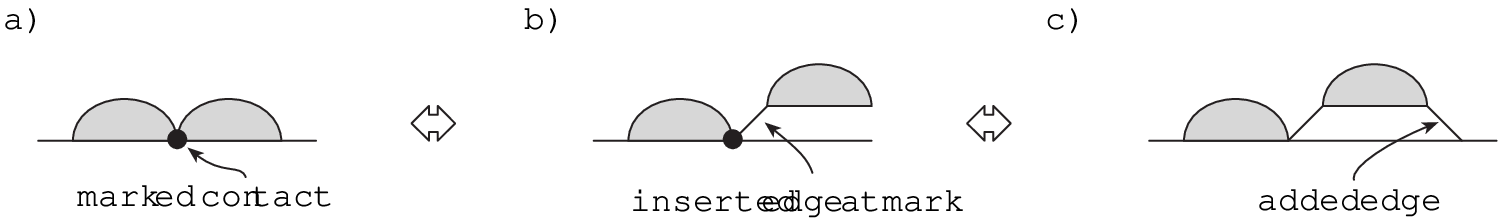}}
\newcommand{\fryingpans}{\includegraphics{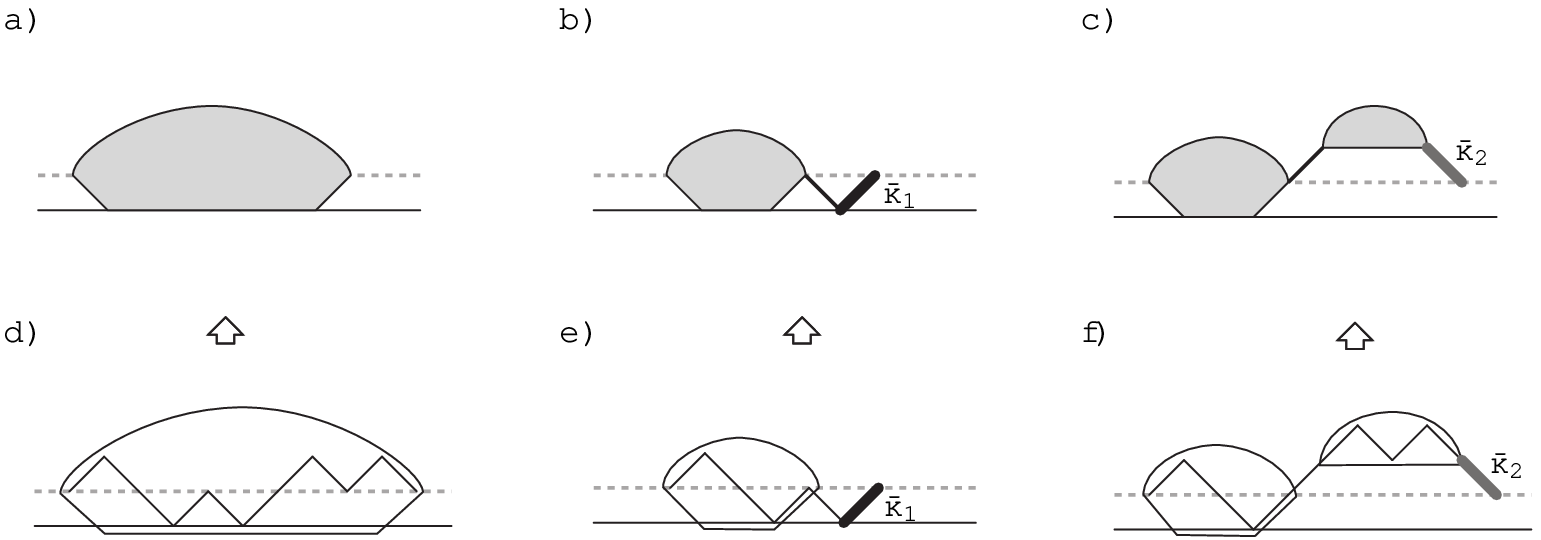}}
\newcommand{\kbseq}{\includegraphics{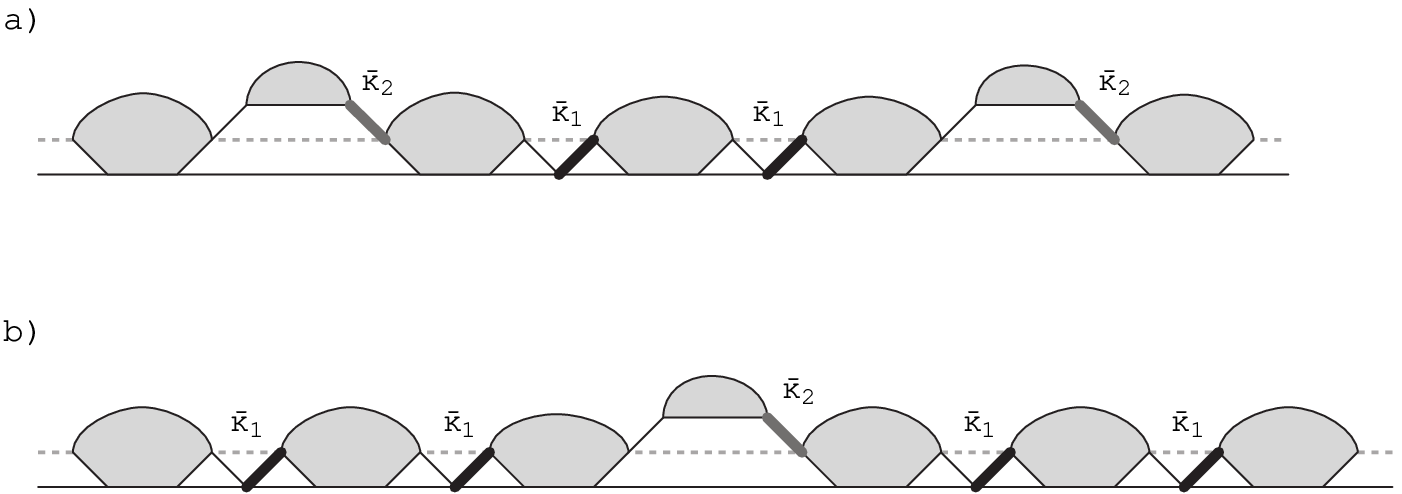}}
\newcommand{\lastfp}{\includegraphics{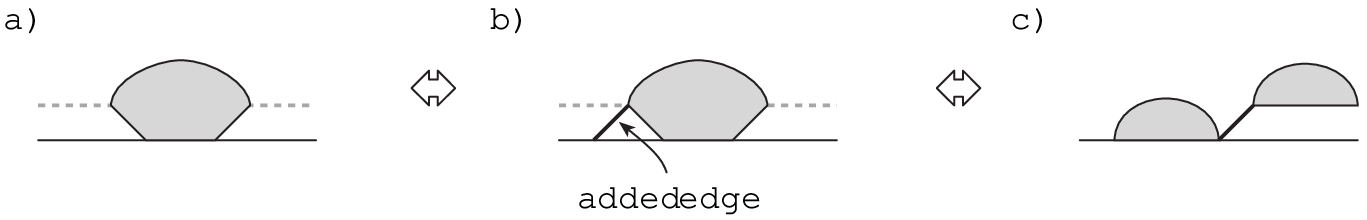}}

\newcommand{\kbo}{\includegraphics{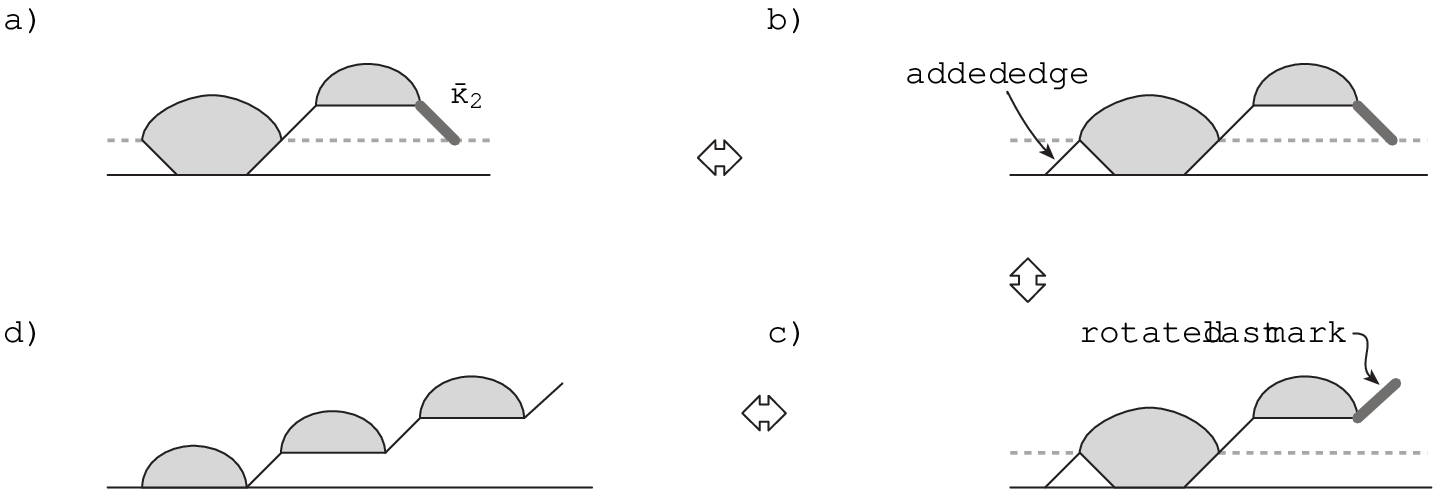}}
\newcommand{\kbt}{\includegraphics{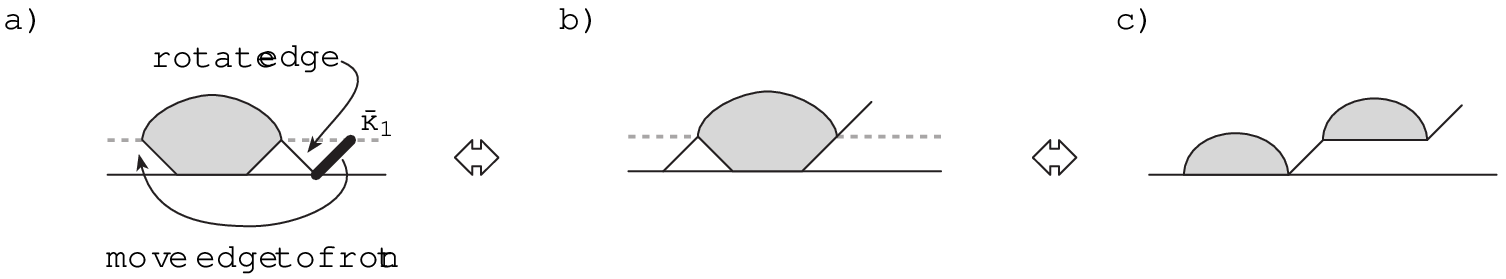}}
\newcommand{\kbokbt}{\includegraphics{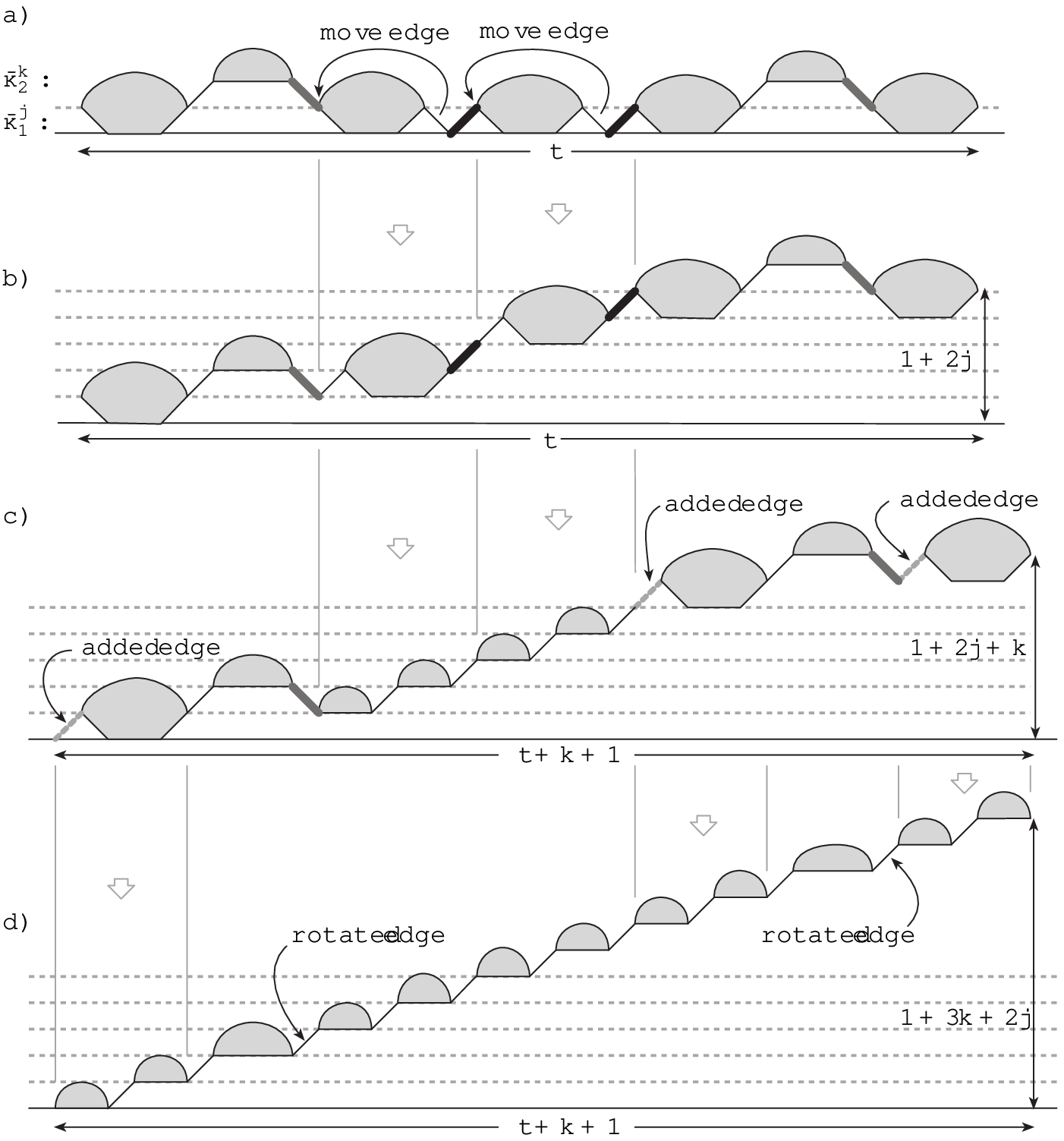}}
\newcommand{\kkseq}{\includegraphics{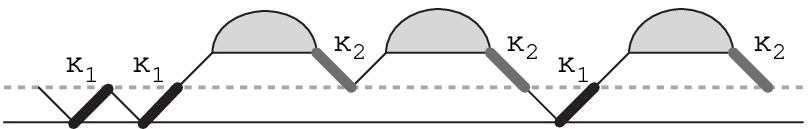}}
\newcommand{\kkbij}{\includegraphics{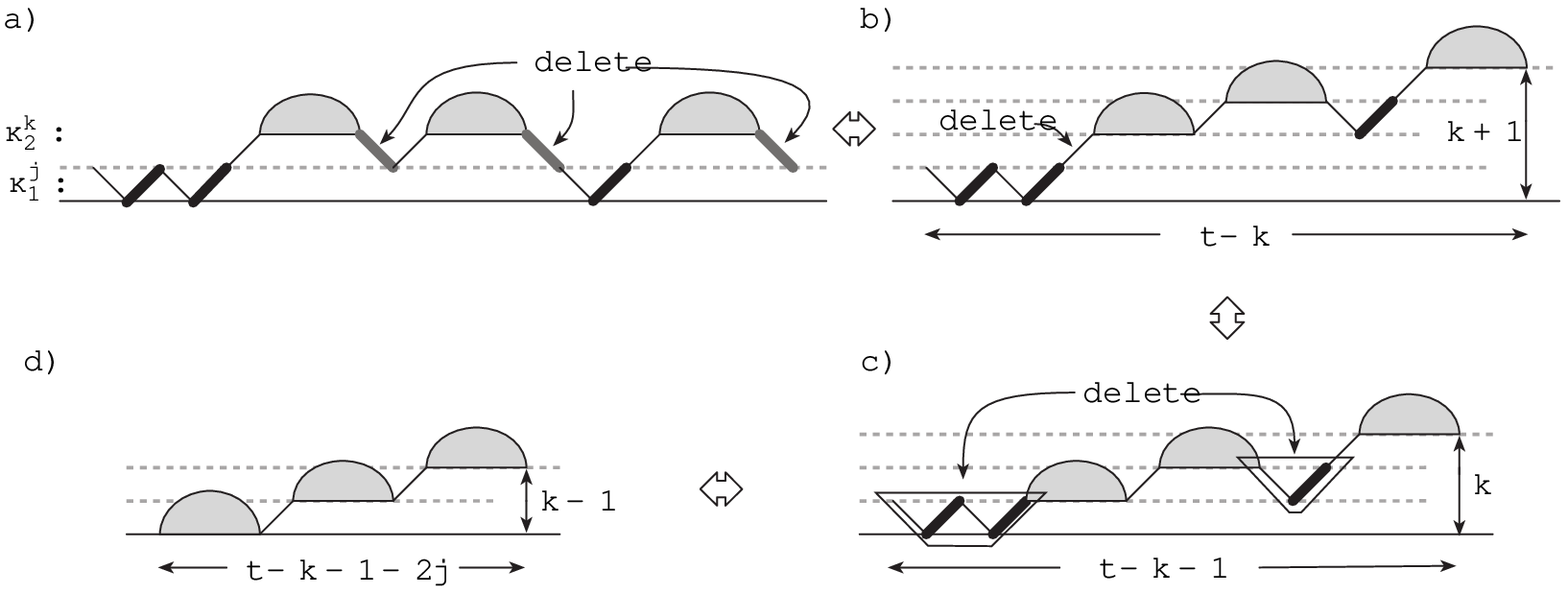}}
\newcommand{\cdseq}{\includegraphics{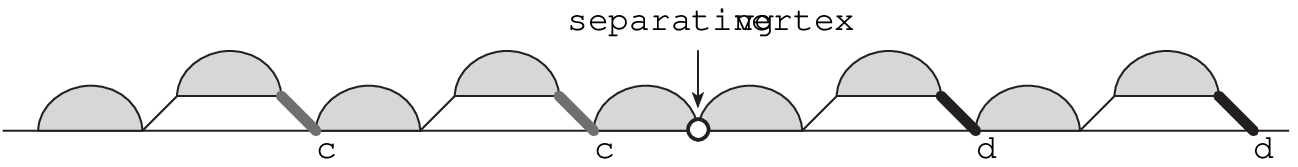}}
\newcommand{\cdbij}{\includegraphics{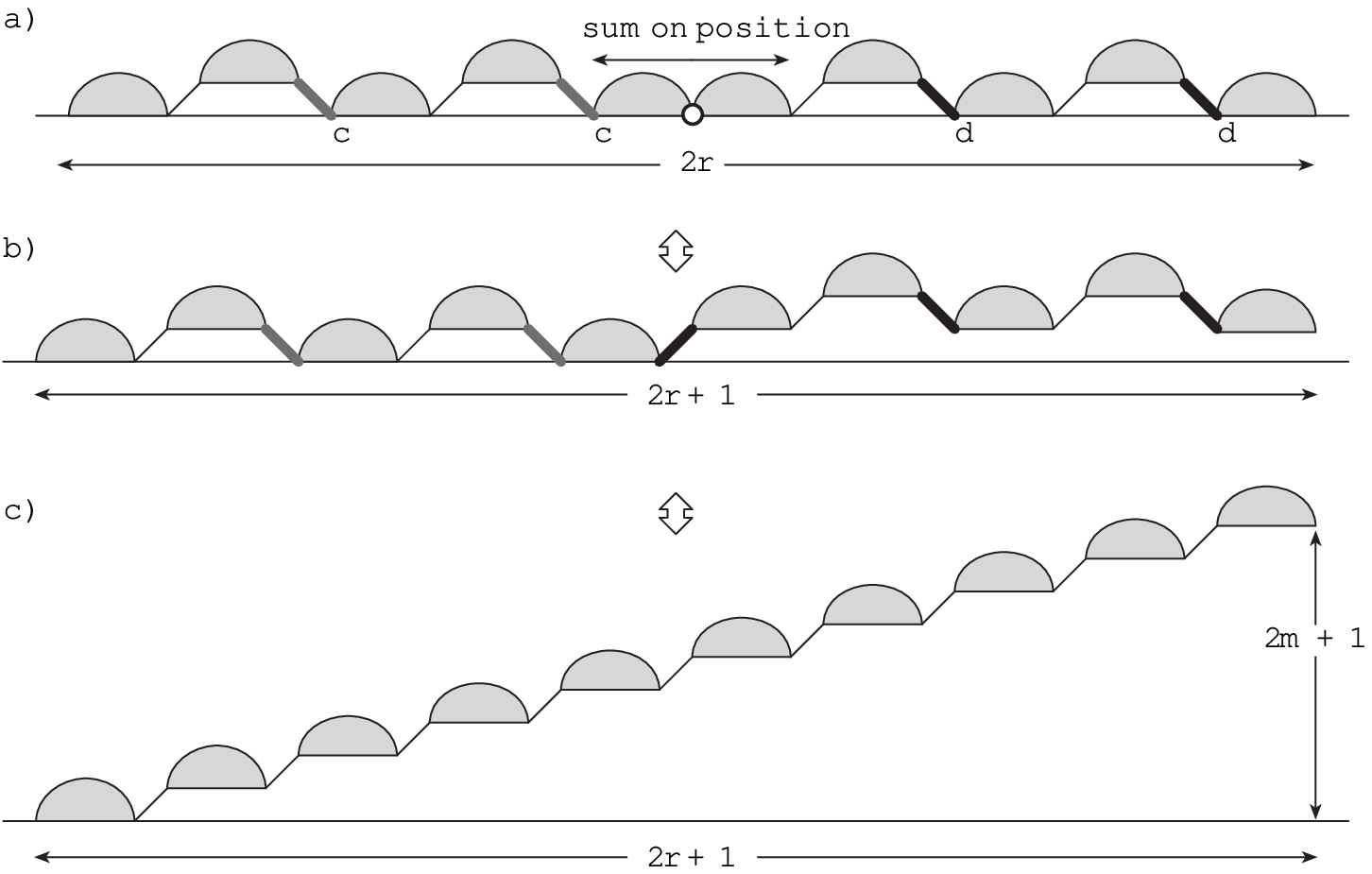}}
\newcommand{\corrpath}{\includegraphics{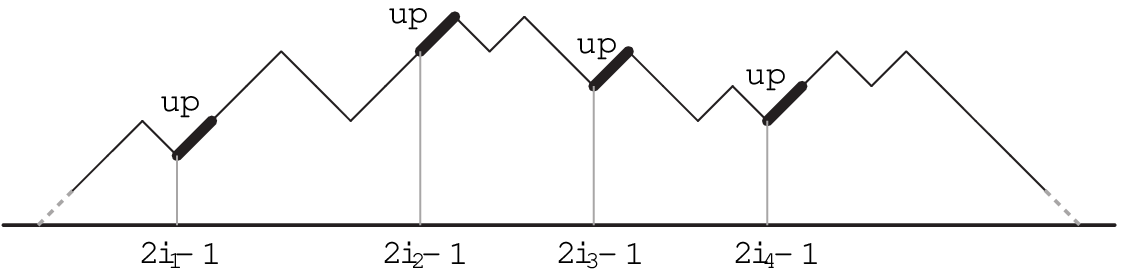}}
\newcommand{\gencor}{\includegraphics{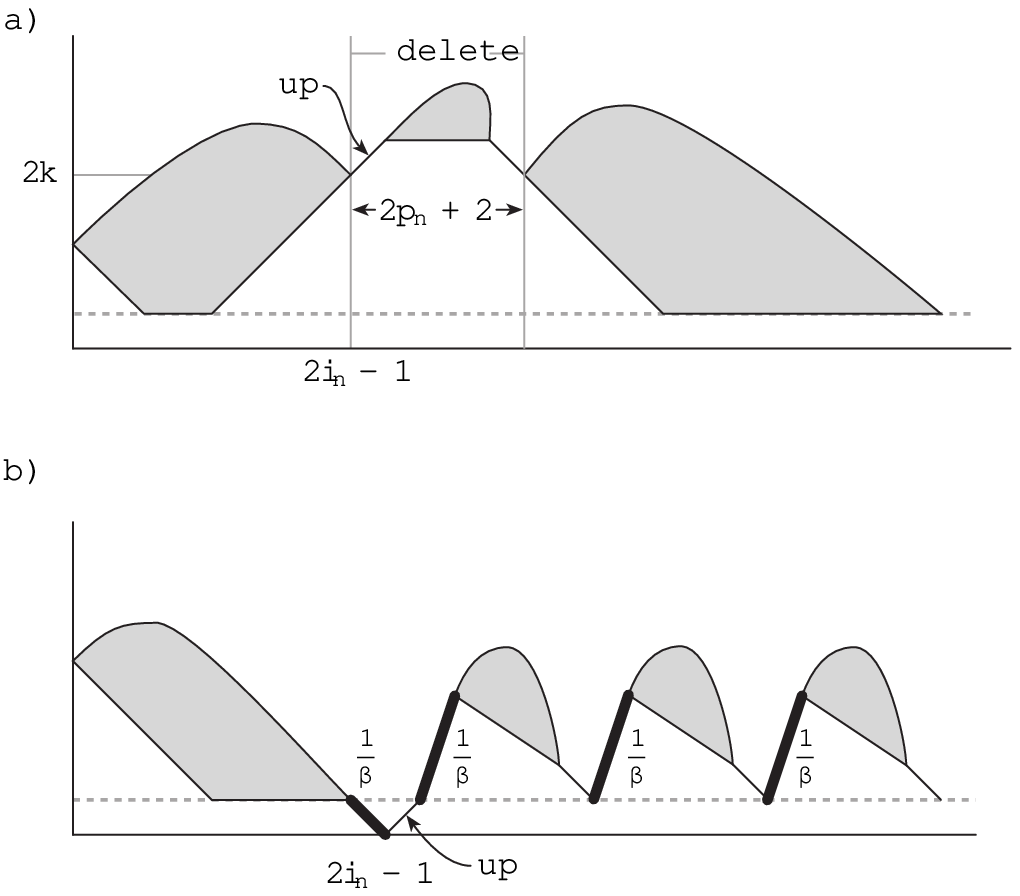}}

 \newcommand{\allPaths}{\includegraphics{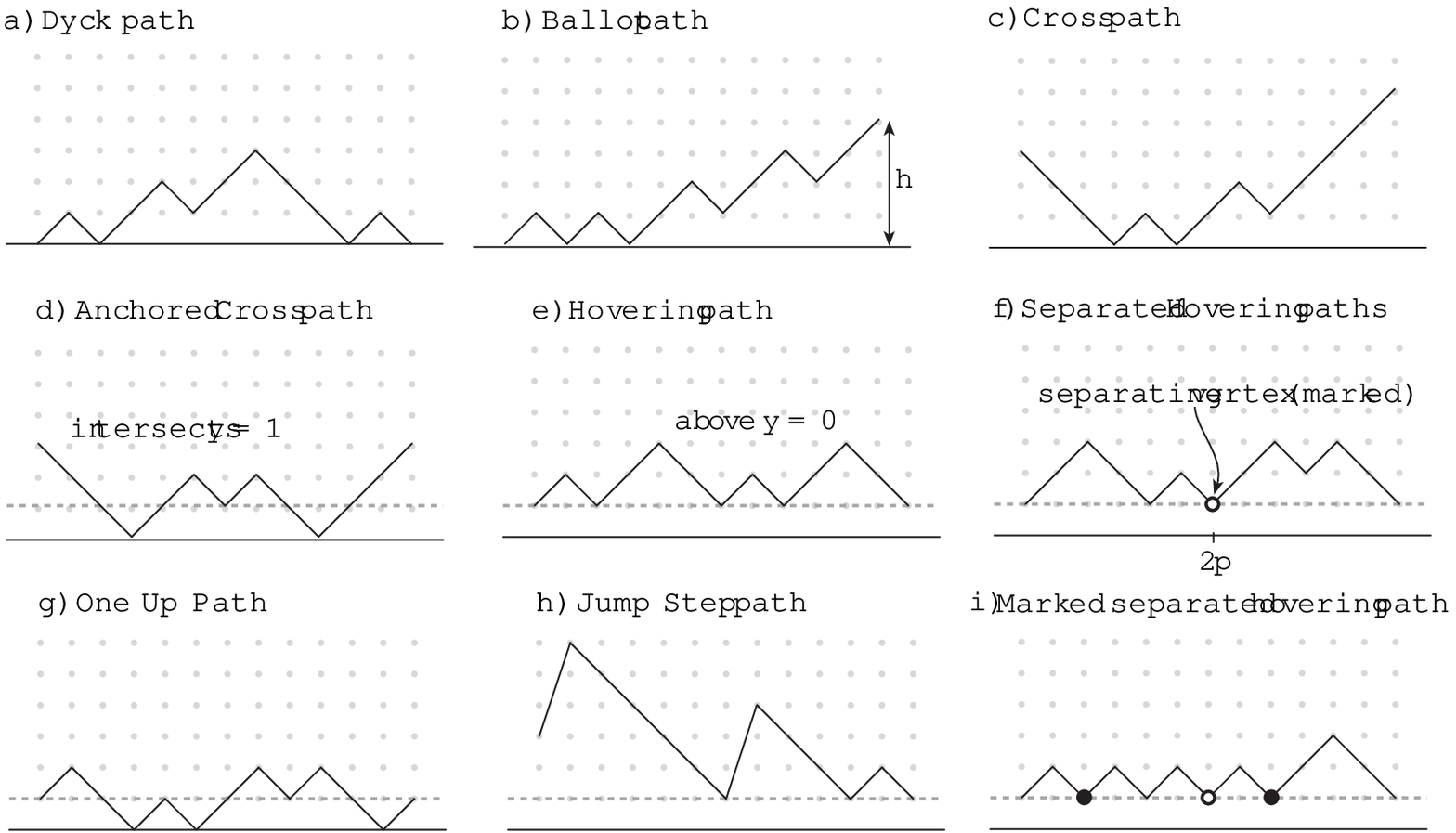}}
 \newcommand{\twocontact}{\includegraphics{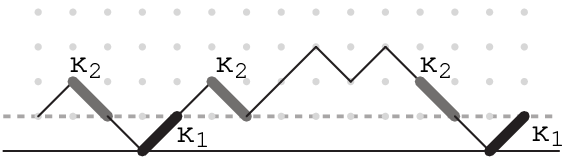}}
\newcommand{\paths}{\includegraphics{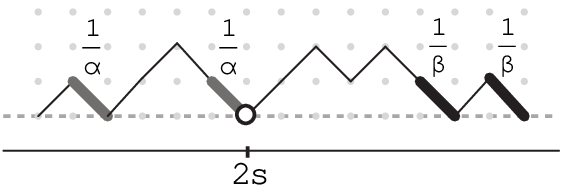}}
\newcommand{\correlation}{\includegraphics{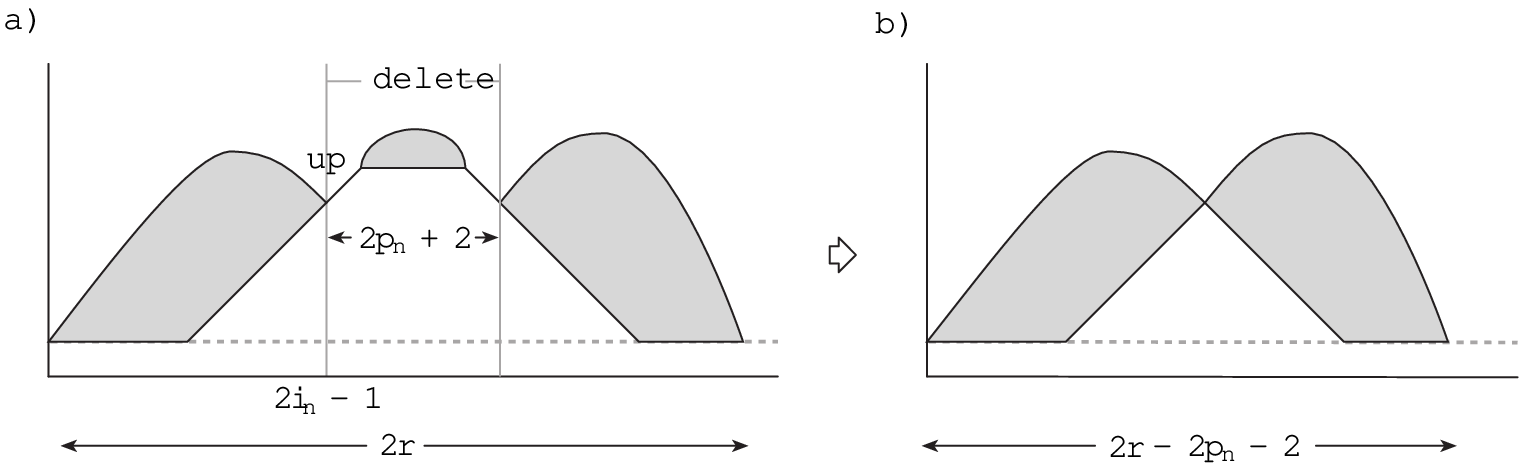}}
\newcommand{\reps}{\includegraphics{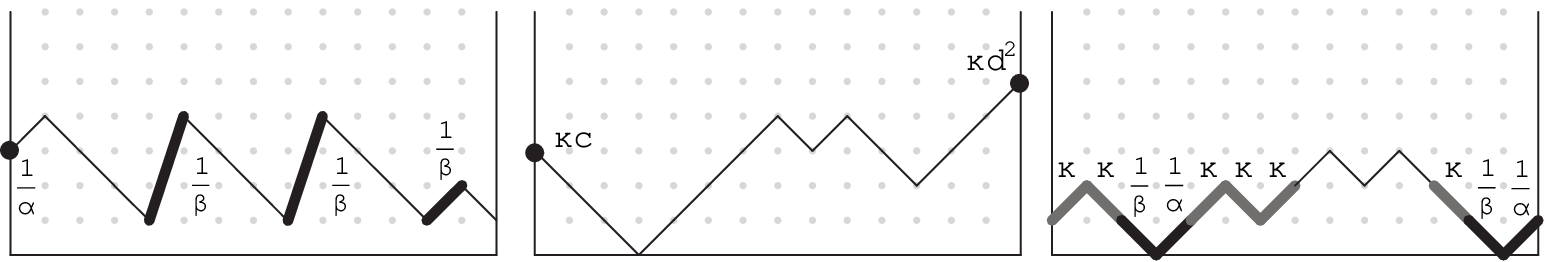}}

\newcommand{\latticeFig}{\includegraphics{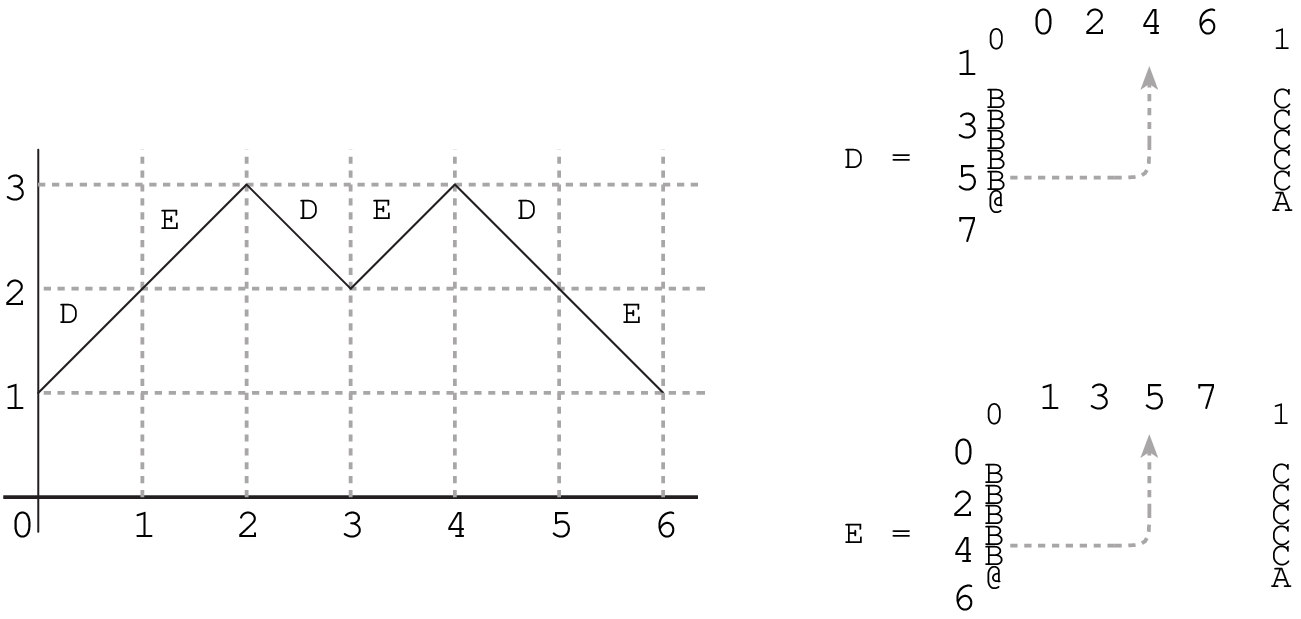}}
\newcommand{\phasediag}{\includegraphics{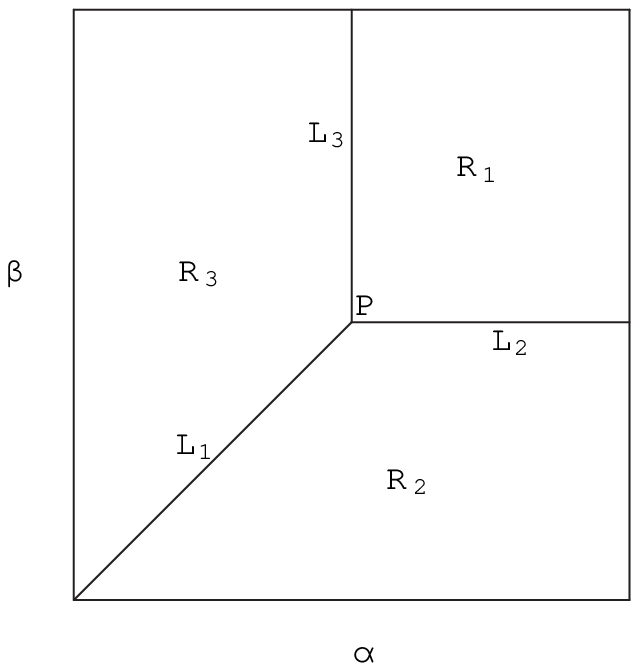}}
\newcommand{\markedBij}{\includegraphics{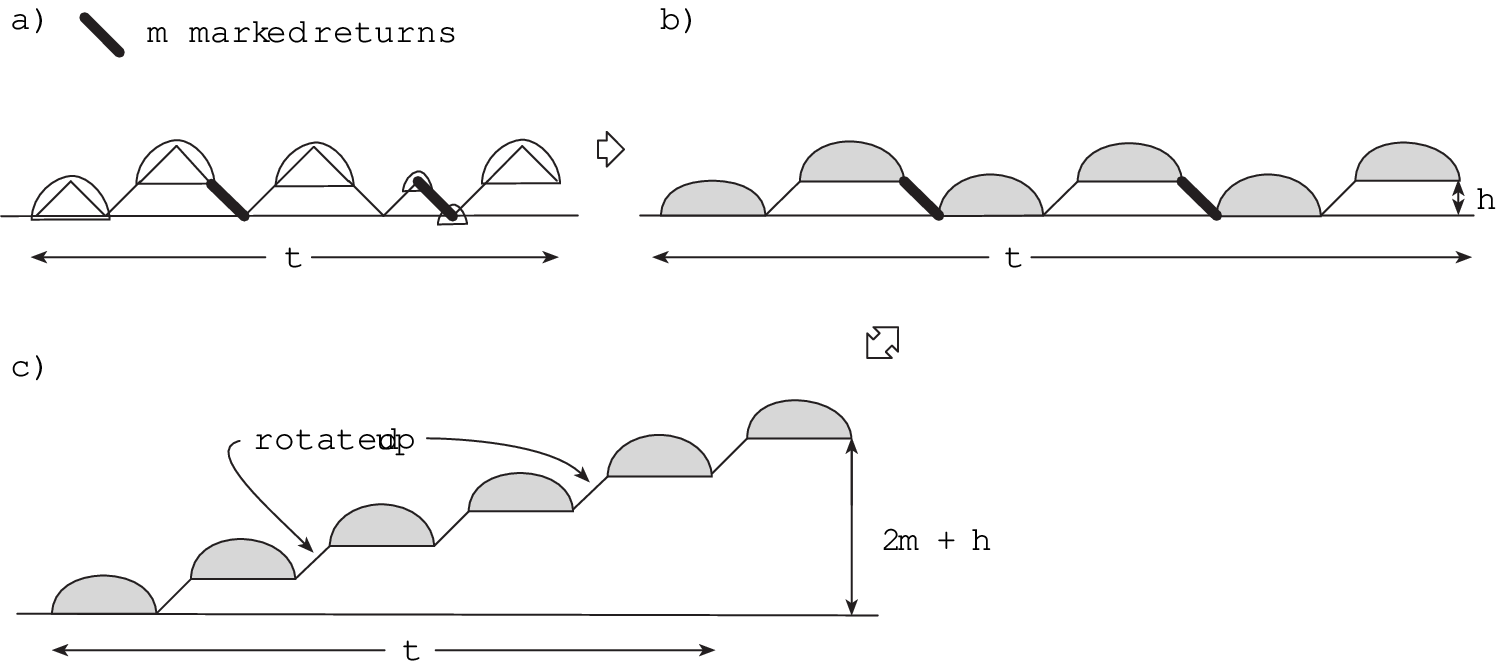}}

\newcommand{\Dyck}{P^{(D)}}
\newcommand{\Ballot}{P^{(B)}}
\newcommand{\Cross}{P^{(C)}}
\newcommand{\OneUp}{P^{(O)}}
\newcommand{\Jump}{P^{(J)}}

\newcommand{\Hovering}{P^{(H)}}
\newcommand{\SepHovering}{P^{(sH)}}

\def\a{\alpha}
\def\b{\beta}
\def\al{\alpha}
\newcommand{\be}{\beta}

\newcommand{\ka}{\kappa}

\def\k{\kappa}

\def\La{\Lambda}
\newcommand{\om}{\omega}
\newcommand{\oc}{\omega_c}
\newcommand{\od}{\omega_d}

\def\ha{\frac 12}

\newcommand{\bb}{\bar{\beta}}
\newcommand{\ab}{\bar{\alpha}}

\newcommand{\ra}{\rangle}
\newcommand{\la}{\langle}

\theoremstyle{plain}

\newtheorem{lemma}{Lemma}
\newtheorem{cor}{Corollary}
\newtheorem{corollary}{Corollary}
\newtheorem{prop}{Proposition}
\newtheorem{defin}{Definition}

\theoremstyle{definition}

\newcommand{\stepSet}{\mathcal{S}}
\newcommand{\edgeseq}{\mathcal{E}}

\newcommand{\integer}{\mathbb{Z}} 
\newcommand{\DirectedSquareLattice}{\Xi}

\newcommand{\pf}{Z}
\def\Za{\pf^a}

\def\half{\frac{1}{2}}

\begin{document} 
%
%
  \DeclareGraphicsExtensions{.eps}
%

%
\pagenumbering{roman}
\setcounter{page}{0}

\title{Asymmetric Exclusion Model and Weighted Lattice Paths.}

\author{R. Brak\dag\ and J. W. Essam\ddag 
        \thanks{email: {\tt r.brak@ms.unimelb.edu.au, 
j.essam@rhul.ac.uk}}
\vspace{0.25 in}\\
         \dag Department of Mathematics and Statistics,\\
         The University of Melbourne\\
         Parkville,  Victoria 3010,\\
         Australia\\
\vspace{0.1 in}\\
         \ddag Department of Mathematics and Statistics,\\
         Royal Holloway College, University of London,\\
         Egham, Surrey TW20 0EX,\\
         England.}

\date{\textbf{\today}} 
 
\maketitle 
 
\begin{abstract} 
We show that the known matrix representations of the stationary state algebra of the
Asymmetric Simple Exclusion Process (ASEP) can be interpreted combinatorially as various weighted lattice paths. 
This interpretation enables us to use the constant term method (CTM) and bijective 
combinatorial methods to express many forms
of the ASEP normalisation factor in terms of Ballot numbers. One particular lattice
path representation shows that the coefficients in the recurrence relation for
the ASEP correlation functions are also Ballot numbers.

Additionally, the CTM has a strong combinatorial connection
which leads to a new ``canonical'' lattice path representation and to the
``$\omega$-expansion'' which provides a uniform approach to computing the
asymptotic behaviour in the various phases of the  ASEP.

The path representations enable the ASEP normalisation factor to be seen 
as  the partition function of a more general polymer chain 
model having a two parameter interaction with a surface.

\end{abstract}

\vspace{1cm} 
 
\noindent{\bf PACS 
numbers:} 05.50.+q, 05.70.fh, 61.41.+e \bigskip

\noindent{\bf Key words: Asymmetric simple exclusion process, lattice paths,
constant term method, enumerative combinatorics } 
\vfill
\pagenumbering{arabic}
\newpage
\section{Background and notation.}
\begin{figure}[ht!]
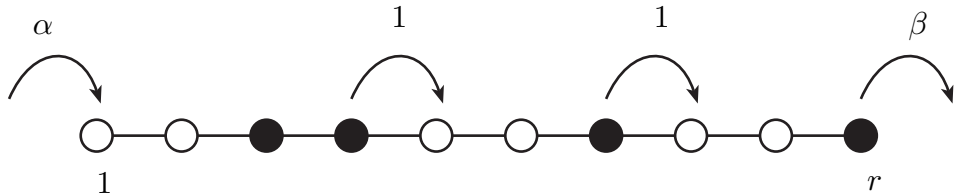

\centering
\asep

\caption{\it The ASEP model. }
\label{fig:asep}
\end{figure}

The Asymmetric Simple Exclusion Process (ASEP) is a simple hard core hopping particle model.
 It consists of a line segment  with $r$ sites. Particles are allowed to hop on to site 1 if it is empty, 
with rate $\alpha$. Any particles on sites $1$ to $r-1$ hop on to a  site to their right  if it is empty, 
with rate $1$. A particle on site $r$ hops off with rate $\beta$ -- as illustrated in figure \ref{fig:asep}. 
The state of the system at any time is defined by the set of indicator variables $\{\tau_1,\cdots,\tau_r\}$, 
where 
\[
\tau_{i}=\begin{cases}
1 & \text{if site  $i$ is occupied}\\
 0 & \text{otherwise}\end{cases} 
\]
and the probability, $P(\vec\tau;s)$ of the system  being in state $\vec\tau =(\tau_1,\cdots,\tau_r)$, at time 
$s$ given some initial state at $s=0$, satisfies a master equation (see \cite{DEHP} for details).

For a review of the extensive literature on this model see Derrida \cite{D}.
A significant step forward in the understanding of the mathematical aspects of the model was made with the 
realization, by Derrida et al \cite{DEHP}, that a stationary state solution, $P_{S}(\vec\tau) $, of the master equation  
could be determined by the matrix product Ansatz
\begin{equation}
P_{S}(\vec\tau) = \la W |\prod_{i=1}^{r}(\tau_{i}D+(1-\tau_{i})E)|V \ra/Z_{2r} \label{prob}
\end{equation}
with normalisation  factor $Z_{2r}$ given by
\begin{equation}
Z_{2r} =\la W |(D E )^{r}|V \ra
\label{eq:norm}
\end{equation}
provided that $D$ and $E$  satisfy the DEHP algebra
\begin{subequations}\label{algebra}
\begin{align}
\label{eqn:de}
    D+E&=DE   
\intertext{where $\la W|$ and $|V\ra$ are the eigenvectors}
    \la W|E&=\frac{1}{\alpha}\la W|,\qquad
     D |V\ra =\frac{1}{\beta }|V\ra.
\end{align}
\end{subequations}
These equations are sufficient to determine $P_{S}(\vec\tau)$ but Derrida et al \cite{DEHP} also gave several
interesting matrix representations of $D$ and $E$ and the vectors $|V \ra$ and $\la W |$,  any one of which may be used
to determine $P_{S}(\vec\tau)$. 
 
The primary result of this paper is to show that each of the three matrix representations of the DEHP algebra 
can be interpreted as a transfer matrix for a different weighted lattice path problem. This then allows the 
normalisation, correlation function and other properties of the ASEP model to be interpreted combinatorially 
as certain weighted lattice path configuration sums -- see section 2. One of the  path connections is similar to that  discussed in Derrida et.\ al.\ \cite{NWA}

The lattice path interpretation has two primary consequences: The first is that it provides  a starting point for a new method (the ``constant term'' or CT method) for calculating the normalisation and correlation functions -- see section 3. This reproduces several existing results (but by a new method) and also  provides several  new results. One of note, the ``$\omega$-expansion'', arises from a  rearrangement of the constant term expression which leads to form of the normalisation in terms of the variables
$\om_c \equiv \a(1-\a)$ and $\om_d \equiv \b(1-\b)$. The coefficients  in this expansion are Catalan numbers, the asymptotic form of which enables a uniform approach to computing the asymptotic behaviour of $Z_{2r}$ as $r\rightarrow \infty$ in the various phases of the  ASEP model -- see section 5. The results agree with those found in \cite{DEHP},  by steepest descent methods.

A bonus following from the lattice path interpretation of the algebra representations is that one of the 
lattice path interpretations  (a slight variation of representation 3) has a natural interpretation as a polymer 
chain having a two parameter $(\k_1,\k_2)$ interaction with a surface. In this context $Z_{2r}$ is a partition function for 
the ``two-contact'' polymer model -- see section 4. We  also obtain recurrence relations (on the length variable) 
for the partition function of this polymer model and hence also for the ASEP normalisation.

The second primary  consequence of the lattice path interpretation follows from the CT method itself, as the  CT method has  very natural combinatorial interpretations. For example,  the normalisation  can be written in several different polynomial forms depending on which variables you use:
\begin{align}
\label{eq:normforms}
Z_{2r}  & =\sum_{n,m}p^{(1)}_{n,m}{\bar\alpha}^{n}{\bar\beta}^{m}  
  & \text{(see -- \eqref{Zabbb})}\\
Z_{2r}  & =\sum_{n,m}p^{(2)}_{n,m} c^{n} d^{m}   
  & \text{(see -- \eqref{Z11cd})}\\
Z_{2r}  & =\sum_{n,m}p^{(3)}_{n,m}\kappa_{1}^{n}\kappa_{2}^{m}   
    & \text{(see -- \eqref{k12})}\\
Z_{2r}  & =\sum_{n,m}p^{(4)}_{n,m}\bar{\kappa}_{1}^{n}\bar{\kappa_{2}}^{m}  
     & \text{(see -- \eqref{Z19})} 
 \end{align}
where all the polynomial coefficients, $p^{(i)}_{n,m}$, are integers.  
(All the variables in these polynomials are related by simple equations eg.\ $\ab = 1/\alpha$, $c=\ab -1$ etc.\ -- 
see above and in later sections).     Since, in each of these cases, the normalisation arises from a weighted 
lattice configuration sum, all the above coefficients have a direct combinatorial interpretation as enumerating 
a particular subset of the paths (eg.\ those with exactly $m$ steps with the first weight  and exactly $n$ steps with the second weight).

However, we show that each of the above polynomial coefficients has an alternative combinatorial interpretation  
which corresponds to  enumerating a different, \emph{unweighted}, set of lattice paths eg.\   in \cite{DEHP} 
the coefficient, $p^{(1)}_{m,n}$ was given as 
\[
p^{(1)}_{n,m-n}  =\frac{m(2r-m-1)!}{r!(r-m)!}
\]
which,  with a simple rearrangement,   can be seen to be a particular ``Ballot number''
\[
p^{(1)}_{n,m-n}=  B_{2r-m-1,m-1}
\]
where $B_{s,h}$  enumerates Ballot paths (see section 3.1 for details) of length $s$ and height $h$. 
Thus $p^{(1)}_{n,m-n}$ which enumerates a special set of paths with $m$ weights of type $\bar\alpha$ 
and $n-m$ weights of type $\bar\beta=1/\beta$ is seen to be determined in terms of
 the much simpler combinatorial problem 
of enumerating  \emph{unweighted}  Ballot paths of length   
$2r-m-1$ and height $m-1$. This correspondence between between the two combinatorial problems 
(one pair for each coefficient) arises as a bijection between the two  path problems. This result may be 
turned around: If a bijection between a particular  Ballot path problem and the weighted path problem 
can be proved then it provides an alternative derivation of the normalisation polynomial.


Finally, in section 6, the recurrence relations for the ASEP correlation functions derived in \cite{DEHP} are shown to follow from the lattice path interpretations.  The coefficients of the terms in the recurrence relation  are also seen to be various Ballot numbers.

\section{Matrix Representations and Lattice Path Transfer 
Matrices.}\label{matrepsec}

Derrida et al  \cite{DEHP}, provided three different matrix representations for the
ASEP algebra. Representation one,
\begin{equation}
D_{1}= 
\begin{pmatrix}\bb&\bb&\bb&\bb&\bb&\cdots\\
0&1&1&1&1&\cdots\\
0&0&1&1&1&\cdots\\
0&0&0&1&1&\cdots\\
\vdots&\vdots&\vdots&\vdots&
\end{pmatrix}  \qquad E_{1}= 
\begin{pmatrix}0&0&0&0&0&\cdots\\
1&0&0&0&0&\cdots\\
0&1&0&0&0&\cdots\\
0&0&1&0&0&\cdots\\
\vdots&\vdots&\vdots&\vdots&
\end{pmatrix}  
\label{eq:rep1}
\end{equation}
\begin{equation}
\la W_{1}|=(1,\ab,\ab^{2},\ab^{3},\ldots)\qquad |V_{1}\ra = (1,0,0,0,\ldots)^{T}
\label{eq:repv1}
\end{equation}
where 
\[ \ab=1/\al,\qquad \bb=1/\be,
\]
representation two,
\begin{equation}
D_{2}= 
\begin{pmatrix}1&1&0&0&0&\cdots\\
0&1&1&0&0&\cdots\\
0&0&1&1&0&\cdots\\
0&0&0&1&1&\cdots\\
\vdots&\vdots&\vdots&\vdots&
\end{pmatrix}  \qquad E_{2}= 
\begin{pmatrix}1&0&0&0&0&\cdots\\
1&1&0&0&0&\cdots\\
0&1&1&0&0&\cdots\\
0&0&1&1&0&\cdots\\
\vdots&\vdots&\vdots&\vdots&
\end{pmatrix}  
\label{eq:rep2}
\end{equation}
\begin{equation}
\la W_{2}|=\ka(1,c,c^{2},c^{3},\ldots)\qquad |V_{2}\ra =
\ka(1,d,d^{2},d^{3},\ldots)^{T}
\label{eq:repv2}
\end{equation}
where  
\[
c=\ab-1,\qquad d=\bb-1,
\]
 and representation three 
\begin{equation}
D_{3}= 
\begin{pmatrix}\bb&\ka&0&0&0&\cdots\\
0&1&1&0&0&\cdots\\
0&0&1&1&0&\cdots\\
0&0&0&1&1&\cdots\\
\vdots&\vdots&\vdots&\vdots&
\end{pmatrix}  \qquad E_{3}= 
\begin{pmatrix}\ab&0&0&0&0&\cdots\\
\ka&1&0&0&0&\cdots\\
0&1&1&0&0&\cdots\\
0&0&1&1&0&\cdots\\
\vdots&\vdots&\vdots&\vdots&
\end{pmatrix}  
\label{eq:rep3}
\end{equation}
\begin{equation}
\la W_{3}|=(1,0,0,0,\ldots)\qquad |V_{3}\ra = (1,0,0,0,\ldots)^{T}
\label{eq:repv3}
\end{equation}
where 
\[
\ka^{2}=\ab+\bb-\ab\bb=1-cd.
\]

Since these matrices and vectors satisfy the algebraic relations \eqref{algebra}, the normalisation factor 
\eqref{eq:norm} for the ASEP model, can be evaluated using any of the
three formulae
\begin{equation}
Z_{2r}^{(j)}=\la W_{j}|(D_{j}E_{j})^{r}|V_{j}\ra,\qquad j=1,2,3.
\label{eq:pf}
\end{equation}

%
\begin{figure}[ht!]
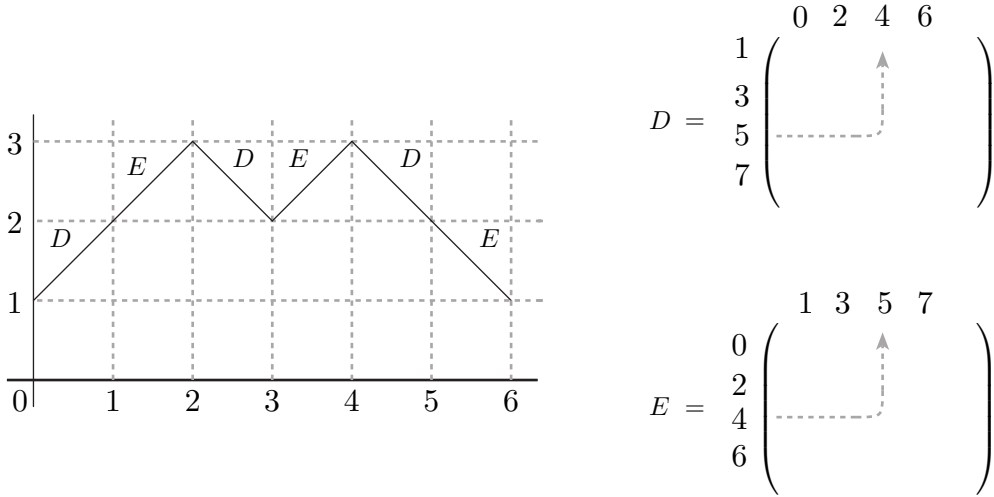

\centering
\latticeFig
\caption{\it a) Action of $D$ and $E$ matrices on the square lattice, b) the
labelling of the matrix elements.}
\label{fig:deac}
\end{figure}

Each of these three representations can be interpreted as the transfer matrix of
a particular weighted lattice path  
problem. If the rows of the $D_j$ matrix and the columns of the $E_j$ matrix are
labelled with odd integers $\integer_{odd} \equiv \{1,3, 5, ...\}$ 
and the columns of $D_j$ and rows of $E_j$ are labelled with even integers
$\integer_{even} \equiv \{0,2,4,...\}$, 
then  $(D_j)_{k,\ell}$ is the weight of a step from an odd height $k$ to even  height
$\ell$ and  $(E_j)_{k,\ell}$ 
is the weight of a step from an even height $j$ to an odd  height $\ell$. Since the
rows and columns are labelled 
with non-negative integers the steps are only in the upper half of
$\mathbb{Z}^2$.

Similarly, the elements of $\la W_j|$  and $|V_j \ra$ are labelled by $\integer_{odd}$
 and are the weights attached to the initial and final vertices of the
paths. 
The matrices $D_j$ and $E_j$ act successively to the left on the initial vector $\la
W_j|$ and $\la W_j|(D_jE_j)^r|V_j \ra$ is the weighted sum over all paths of length $2r$
which begin and end at odd height above the $x-$axis.

\begin{figure}[ht!]
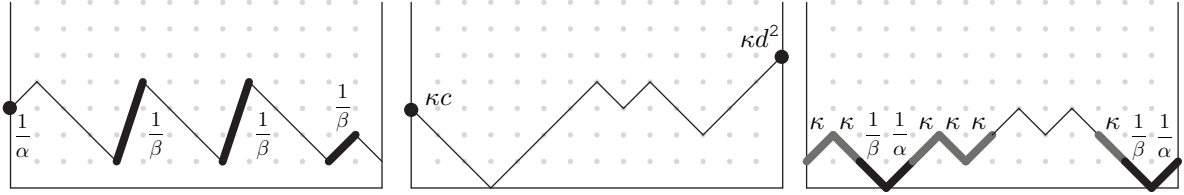

\centering
\reps
\caption{\it Examples of the paths for each of the three representations a)
representation one, ``Jump step paths'', b) representation two, ``Cross paths''
and c) representation three, ``One up paths''. }
\label{fig:repEx}
\end{figure}

An example path for each of the three representations is shown in figure
\ref{fig:repEx}. 
Notice that for the first representation the paths with non-zero weight begin at
any odd height and end at unit height, for the second they both begin and end at
any odd height and 
for the third they begin and end at unit height. All these paths are defined
explicitly below. 
\\

Derrida et.\ al.\ gave another  expression for the normalisation factor
(see equation (39) of \cite{DEHP}). This form does not arise directly from any of
the above three representations, 
however, we will show (corollary \ref{alphabeta}) that it is the partition function $Z_{2r}^{(5)}$, defined in
\eqref{eq:4422}, corresponding to a
``canonical'' path representation (see figure \ref{fig:canp} for an example). In \cite{BER} we provide a 
combinatorial derivation of the equivalence of the above three and a number of other path representations
of the normalisation factor to the ``canonical'' representation.

\subsection{The Lattice Path Definitions.}

We consider  paths whose steps are between the vertices of the half plane   square lattice
$\DirectedSquareLattice= \{(x,y)|\,x\in 
\integer, y
\in \integer^{+}\}$, where $\integer$ (resp.
$\integer^{+}$) is the set of  integers 
(resp. non-negative integers).

\begin{defin}[Lattice paths] A \textbf{lattice path}, $\om$, of length $t\ge 0$
is a sequence of vertices $(v_{0},v_{1},\ldots,v_{t})$ where  $v_{i}\equiv (x_i,y_i)\in
\DirectedSquareLattice$, and for $t>0$, $v_{i}-v_{i-1}\in\stepSet_{i}^{p}$ where
$\{\stepSet_{i}^{p},i=1,\ldots, t\}$ is the set of allowed steps.  For a
particular path, $\om$, denote the corresponding sequence of steps by
$\edgeseq(\om)=e_{1}e_{2}\ldots e_{t}$, 
$e_{i}=(v_{i-1},v_{i})$.   A \textbf{subpath} of length $k$ of a lattice path, $\om$, is a
path defined by a subsequence of adjacent vertices,
$(v_{i},v_{i+1},\ldots,v_{i+k-1}, v_{i+k})$, of $\om$.  
\end{defin}
\begin{figure}[ht!]
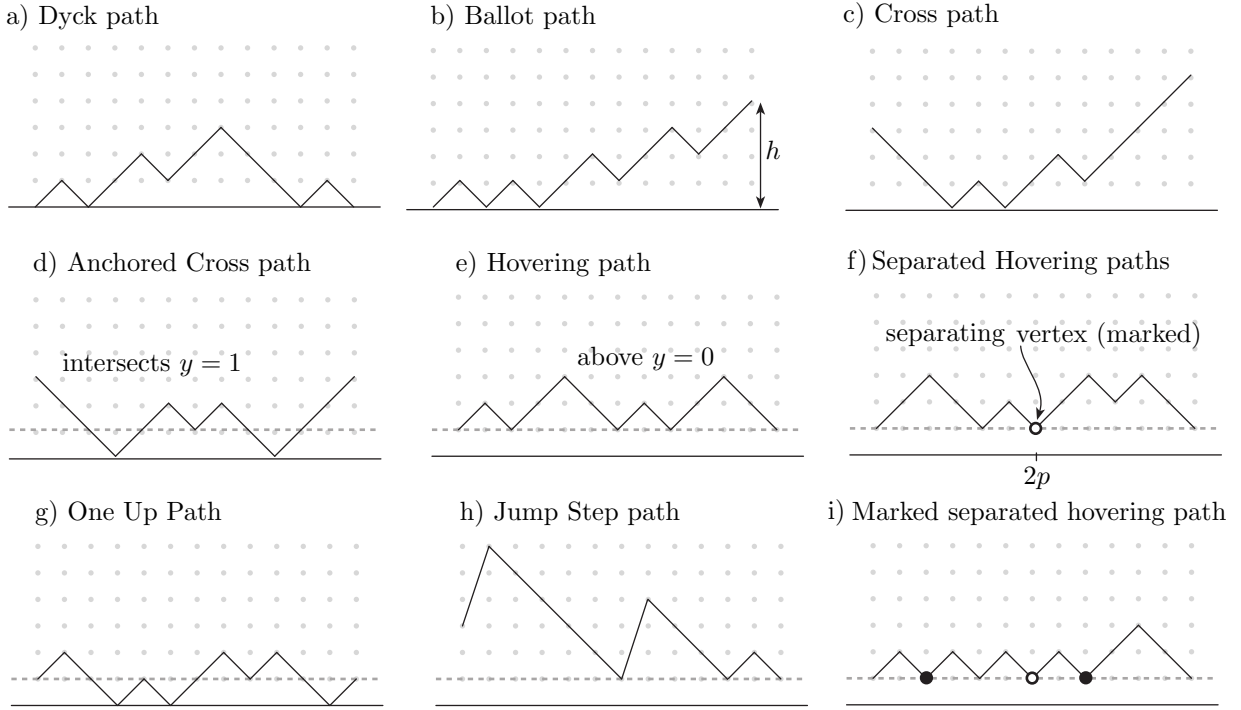

\centering
\allPaths
\caption{\it An example of a a) a Dyck path, b) a Ballot path, c) a Cross path, d) an 
Anchored Cross path, e) a Hovering path, f) a Separated Hovering path, g) a One Up path, 
h) a Jump Step path and i) a Marked separated hovering path.}
\label{fig:allpaths}
\end{figure}

 \begin{defin}[Dyck $\Dyck_{2r}$,  Ballot $\Ballot_{t;h}$,   \& Cross $\Cross_{t;h_1,h_2}$  paths] The set of length $2r$ \textbf{Dyck}
paths, $\Dyck_{2r}$;  length $t$ and height $h$ \textbf{Ballot} paths $\Ballot_{t;h}$,
and  length $t$ and heights $h_1$ and $h_2$\textbf{ Cross}  paths
$\Cross_{t;h_1,h_2}$ all
have step set $\stepSet_{i}^{D}=\{(1,1),(1,-1)\}$.  Dyck paths  have
$v_{0}=(0,0)$ and $v_{2r}=(0,0)$,   Ballot paths have $v_{0}=(0,0)$ and
$v_{t}=(t,h)$   and Cross paths  
have $v_{0}=(0,h_1)$ and $v_{t}=(t,h_2)$, 
$h_1,h_2\ge0$. Denote, $\Cross_{2r}=\bigcup_{h_1,h_2\in \integer_{odd}}
\Cross_{2r;h_1,h_2}$ 
 \end{defin}
\begin{defin}[Elevated Dyck path  (Bubble)]\label{defn:bub} An  \textbf{elevated Dyck path}  or  \textbf{Bubble},  is a
subpath, $ (v_{i}, \ldots , v_{i+k})$, $k\ge0$, 
  for which $y_i=y_{i+k}$ and $y_j\ge y_i$ for
all $i<j<i+k$, $k>0$.  If $k=0$ then the elevated Dyck path is a single vertex.
 \end{defin}
\begin{defin}[Anchored Cross $P^{(aC)}_{t;h_{1},h_{2}}$ paths] The subset $P^{(aC)}_{t;h_{1},h_{2}}\subset
P^{(C)}_{t;h_{1},h_{2}}$ of \textbf{Anchored Cross Paths} is defined as all the
paths in $P^{(C)}_{t;h_{1},h_{2}}$ with at least one vertex in common with the
line $y=1$. Denote, $P^{(aC)}_{2r}=\bigcup_{h_1,h_2\in \integer_{odd}}
P^{(aC)}_{2r;h_1,h_2}$ 
\end{defin}
\begin{defin}[One Up $\OneUp_{2r}$  paths] The set of  \textbf{One Up} paths, $\OneUp_{2r}$
is the set of  lattice paths of length $2r$ which have step set $\stepSet_{i}^{D}=\{(1,1),(1,-1)\}$, 
$v_{0}=(0,1)$ and $v_{2r}=(2r,1)$, i.e. $\OneUp_{2r}=\Cross_{2r;1,1}$.
\end{defin} 
\begin{defin}[Jump Step $\Jump_{t;h}$ paths]\label{def:js} The set of \textbf{Jump Step} paths of length $t$,
$\Jump_{t;h}$, is the set of  lattice paths which have step set 
\begin{equation*}
\label{eq:jspst }
\stepSet_{i}^{J}=\left\{ 
\begin{array}{ ll}
   \{(1,-1)\}   & \text{for $i$ even}  \\
   \{(1,-1)\}\cup\{(1,2\ell+1) | \ell\in \integer^{+}  \}   &    \text{for $i$ odd}
\end{array}\right.
\end{equation*} 
with  $v_{0}=(0,h)$, $h \in \integer_{odd}$, $v_{t}=(t,1)$ . The ``height'',
 $g_i$,  of   a step $e_i = (v_{i-1},v_{i})$ is defined as  
$y_{i}-y_{i-1}$.   Odd steps with  $g_i \in \{3,5,\dots\}$ will be  called ``jump'' steps. If $g_i=1$ (resp.
$g_i=-1$) the step is called an ``up'' (resp. ``down'') step.
Denote, $\Jump_{2r}=\bigcup_{h\in \integer_{odd}} \Jump_{2r;h}$
\end{defin}
{\it Note:}\,  Jump step paths never visit the $x$-axis since such a visit can only occur on an odd step
and return to $y=1$ is impossible since all even steps are down. 

\begin{defin}[Hovering $\Hovering_{2r}$,   Separated Hovering $\SepHovering_{2r;2p}$  and Marked Separated Hovering $ P^{(mH)}_{2r;2p}$ paths] \label{def:mh}
\textbf{Hovering} paths, $\Hovering_{2r}$ are  One Up paths with no vertex on the $x$-axis. \textbf{Separated Hovering}
paths, $\SepHovering_{2r;2p}$, $0\le p \le r$, are the subset of the Hovering paths
which have $v_{2p}=(2p,1)$. The vertex $v_{2p}$ is marked (with an empty circle -- see figure \ref{fig:allpaths}f ) and known as the \textbf{separating vertex}. \textbf{Marked  separated hovering} paths, $ P^{(mH)}_{2r;2p}$, are obtained from $\SepHovering_{2r;2p}$ by marking (with an solid circle -- see figure \ref{fig:allpaths}i ) subsets of the steps (or vertices) which return to $y=1$.
\end{defin}
{\it Note:} For all paths considered 
$x_i+y_i$ is either odd for all $i$ or even for all $i$, i.e. the paths are confined either to the
odd sublattice or the even sublattice.

\begin{defin}[Contacts, Returns]\label{defn:ret} A vertex of a Ballot or Dyck path in common with the
$x$-axis is called a {\textbf{contact}}. All contacts except the initial one are
called  {\textbf{returns}}.
The polynomial $R_t(h;\ka) = \sum_{\om \in P_{t;h}^{(B)}} \ka^{\rho(\om)}$, where $\rho(\om)$ is 
the number of returns
for the path $\om$, is called the {\textbf{return polynomial}} for Ballot paths.
\end{defin}

\subsection{Weights and lattice path representations.}

For each of the three different representations the following lemma converts the
matrix formula \eqref{eq:pf} for the normalization
factor into a sum over one of the path sets defined above, where the summand is
a product of  the step weights $w^{step}_{j}(e_{i})$,   an initial  vertex weight $w^{i}_{j}(v_{0})$, and a  final vertex weight  $w^{f}_{j}(v_{t})$.

\begin{lemma} The normalisation factor, $Z_{2r}^{(j)}$ for each of the three
matrix representations, $j=1,2,3$, can be written as
\begin{align} Z_{2r}^{(j)}&=\sum_{\om\in 
P^{(j)}_{2r}}W^{(j)}(\om)\label{eq:448}\\
\intertext{with} 
    W^{(j)}(\om)&=w^{i}_{j}(v_0)\left[
\prod_{i=1}^{2r} w^{step}_{j}(e_{i})
\right]w^{f}_{j}(v_{2r})\label{eq:449}
\end{align}


 \noindent where $P^{(1)}_{2r} =P^{(J)}_{2r}$, $P^{(2)}_{2r} =P^{(C)}_{2r}$, and
 $P^{(3)}_{2r} =P^{(O)}_{2r}$, and the weight $W^{(j)}(\om)$ of a particular path
 $\om$ with step sequence, $\edgeseq(\om) = e_{1}\ldots
 e_{2r}$ is defined, for each of the three cases, as follows.
 
\noindent For $j=1$
\begin{subequations}\label{jweights}
\begin{align} w^{i}_{1}\bigl((0,2k+1)\bigr)&=\ab^{k}\qquad k\in \integer^{+}\\
w^{step}_{1}(e_{i})&=
\begin{cases}
\bb & \text{if 
$e_{i}=\bigl((i-1,1),(i,2k)\bigr)$, 
$k\in \integer^{+}$ and 
$i$   odd}\\
1 & \text{otherwise}
\end{cases} \\
w^{f}_{1}(v_{2r})&=1,
\end{align}
\end{subequations}
for $j=2$
\begin{subequations}
\begin{align}
  w^{i}_{2}\bigl((0,2k+1)\bigr)&=\ka\, c^{k}\qquad k\in 
\integer^{+}\\
w^{step}_{2}(e_{i})&=1 \\
w^{f}_{2}\bigl((0,2k+1)\bigr)&=\ka\, d^{k}
\qquad k\in \integer^{+}
\end{align}
\end{subequations}
and for $j=3$
\begin{subequations}\label{eqs:w2} 
\begin{align}
  w^{i}_{3}((0,1))&=1\\
w^{step}_{3}(e_{i})&=
\begin{cases}
\ka & \text{if $e_{i}=\bigl((i-1,1),(i,2)\bigr)$ or
$e_{i}=\bigl((i-1,2),(i,1)\bigr)$}\\
\bb & \text{if 
$e_{i}=\bigl((i-1,1),(i,0)\bigr)$ }\\
\ab & \text{if 
$e_{i}=\bigl((i-1,0),(i,1)\bigr)$ }\\
1 & \text{otherwise}
\end{cases} \\
w^{f}_{3}((0,1))&=1
\end{align}
\end{subequations}
\end{lemma}
\begin{proof} The above lemma is a direct consequence of the lattice path
interpretation of the $D_j$ and $E_j$ matrices as transfer matrices. 
\end{proof}

The equivalence
of the different expressions is a consequence of the invariance of the
normalisation factor under similarity transformations relating the different
matrix representations of $D$ and $E$. 


In order to compute $Z_{2r}^{(3)}$, it turns out to be a little more convenient
to rearrange the weights associated with representation three. If we do so we
obtain the following corollary.
\begin{figure}[ht!]
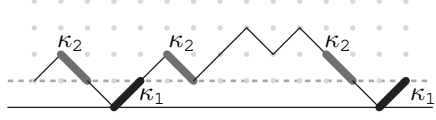

\centering
\twocontact
\caption{\it An example of a representation three, one up path, with the weights
re-organised}
\label{fig:sw}
\end{figure}

\begin{corollary} An equivalent set of weights for $Z_{2r}^{(3)}$, is
\begin{align}\label{w4} w^{i}_{4}((0,1))&=1\\
w^{step}_{4}(e_{i})&=
\begin{cases}
\ka_1= \ab\bb & \text{if $e_{i}=\bigl((i-1,0),(i,1)\bigr)$ }\\
\ka_2=\ka^{2} &   \text{if $e_{i}=\bigl((i-1,2),(i,1)\bigr)$}\\
1 & \text{otherwise}
\end{cases} \\
w^{f}_{4}((0,1))&=1
\end{align} 
\end{corollary}

{\it Note:} this rearrangement of the weights is only valid for computing
$Z_{2r}^{(3)}$. For correlation functions one has less freedom in the weight
rearrangement.

 \begin{figure}[ht!]
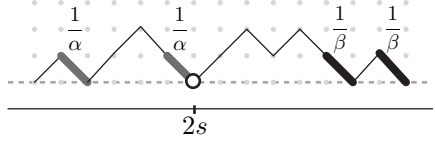

\centering
\paths
\caption{\it An example of a canonical path -- a Separated Hovering path with a
$W^{(5)}$ weighting. }
\label{fig:canp}
\end{figure}
 
Later in the paper (corollary \ref{alphabeta}) we will show that  the 
normalisation factor can also be expressed in terms of Separated Hovering paths.  
\begin{equation} 
Z_{2r}^{(5)}=\sum_{p=0}^r\,\, \sum_{\om_p\in P^{(sH)}_{2r;2p} }
W^{(5)}(\om_p)\label{eq:4422}
\end{equation}
where $W^{(5)}(\om_p)$ is defined by \eqref{eq:449} with 
\begin{subequations}\label{eqs:w5}
\begin{align} w^{i}_{5}((0,1))&=1\\
w^{step}_{5}(e_{i})&=
 \begin{cases}
  \ab & \text{if  $e_{i}=\bigl((i-1,2),(i,1)\bigr)$ and $i\le 2p$}\\
  \bb &   \text{if  $e_{i}=\bigl((i-1,2),(i,1)\bigr)$ and $i> 2p$}\\
  1 & \text{otherwise}
 \end{cases} \\
w^{f}_{5}((0,1))&=1
\end{align}
\end{subequations}
Thus, for any particular path, $\om_{p}$, all the $\ab$ weighted steps (if any)
occur to 
the left of 
vertex $(2p,1)$ and all the $\bb$ weighted steps (if any) occur to the right of 
$(2p,1)$. We call this combination of paths and weights the ``canonical'' path
representation of the normalisation factor. An example is shown in figure \ref{fig:canp}.

\begin{lemma}
Let $P^{(mH)}_{2r;2p}$ be the set of marked  separated hovering paths obtained from $\SepHovering_{2r;2p}$ by marking
subsets of the steps which return to $y=1$ then 
 
\begin{equation}\label{Hsum} 
Z_{2r}^{(2)} = \sum_{p=0}^r\sum_{\om_p \in P^{(mH)}_{2r;2p}}W^{(2a)}(\om_p)
\end{equation}
where the weight $W^{(2a)}(\om_p)$ has a factor $c$ for each marked return step which
occurs to the the left of $v_{2p}$ and a factor $d$ for the other marked return steps. 
\end{lemma}
The lemma is proved in \cite{BER} in two stages. First an involution on $P^{(C)}_{2r}$ for which
$P^{(aC)}_{2r}$ is the fixed set and then a bijection between $P^{(aC)}_{2r}$ and $\bigcup_{p=0}^r P^{(mH)}_{2r;2p}$.

\begin{cor}\label{Z2Z5}
$$Z_{2r}^{(2)} = \left . Z_{2r}^{(5)}\right |_{\ab = 1+c, \bb = 1 +d}$$
\end{cor}
\begin{proof}
Substituting $\ab = 1+c, \,\bb = 1 +d$ in the weight attached to a path $\om_p\in \SepHovering_{2r;2p}$
and expanding leads to a sum of terms obtained by weighting each return step to the right of $v_{2p}$
(if any) with either $1$ or $d$ and the remaining returns (if any) by $1$ or $c$. Marking the subsets of steps 
weighted $c$ or $d$ determines a set of paths belonging to $P^{(mH)}_{2r;2p}$.
\end{proof}

\section{Methods.}
In this section we briefly review methods to be used and results obtained previously \cite{BE} as they
will be required in the next section. One additional new lemma is stated.

\subsection{The constant term method.}

\begin{defin} The constant term operation, $CT[\cdot ]$,  is defined by
\[
CT[f(z)] = \text{ the constant term in the Laurent expansion of  $ f(z)$ about
$z=0$.}
\]
\end{defin}

The number of $t-$step lattice paths with step set $S_i^D$ which begin at $(0,0)$ and end at $(t,y)$ with no further
constraint (i.e. replacing the constraint $y\in \integer^+$ in the definition of $\DirectedSquareLattice$  
by $y\in \integer$) is the binomial coefficient ${t\choose \frac 12(t-y)}$ for which the constant term formula is
\begin{equation}
{t\choose \frac 12(t-y)} = CT[(z + 1/z)^t z^y].
\end{equation}

By the reflection principle \cite{andre}, the number of $t-$step Ballot paths of height $h$ is obtained
by subtracting the number of unrestricted paths which begin at $(0,-2)$ from those beginning at $(0,0)$,
both ending at $(t,h)$.
\begin{equation}\label{CTBallot}
B_{t,h} \equiv |P^{(B)}_{t;h}| = CT[(z + 1/z)^t z^h] - CT[(z + 1/z)^t z^{h+2}] = CT[\La^t z^h (1-z^2)]
\end{equation}
where $\La = z + 1/z$. Differencing the binomial coefficients expresses $B_{t,h}$ in terms of factorials;
for $t+h$ even
\begin{equation}
\label{eq:balnum}
 B_{t,h} =\frac{(h+1)t!}{(\frac{1}{2}(t+h)+1)!(\frac{1}{2}(t-h) )!}.
\end{equation} 
Dyck paths are Ballot paths ending at $(t,0)$ so the number Dyck paths with $2r$ steps is
\begin{equation}\label{CTCat}
|P^{(D)}_{2r}| = CT[\La^t (1-z^2)] = \frac 1{r+1} {2r\choose r} = C_r,
\end{equation}
a Catalan number.

Instead of using the reflection principle the Ballot numbers may be obtained as the solution of the equations, $t,h\ge 1$,
\begin{subequations}
\begin{align}B_{0,h} &= \delta_{h,0}\\
B_{t,0} &= B_{t-1,1}\\
B_{t,h} &= B_{t-1,h-1} + B_{t-1,h+1},  
\end{align}
\end{subequations}   all of which are solved by $CT[\La^t z^h (1-z^2)]$. 

Historically, Ballot numbers arise in the combinatorial problem  of a two candidate  election. If you ask how many ways can $t$ votes be caste such that the first candidate ends $h$ votes ahead of the second candidate and at any stage of the voting never has fewer votes that the second candidate.

\subsection{Partition function for the one contact model.}

Previously \cite{BE}   we proved the following proposition concerning the return polynomial (see definition \ref{defn:ret}).
\begin{prop}  The return polynomial for Ballot paths of length $t$ and height $h$ is given by 
\begin{equation}\label{eq:rpctf}
R_t(h;\ka) = CT[\frac{\La^t z^h (1-z^2)}{1- (\ka-1)z^2}]
\end{equation}
\end{prop}
\begin{proof}
For $t,h\ge 1$, $R_t(h;\ka)$ is determined by the recurrence relations
\begin{subequations}
\begin{align}
\label{eq:rprr}
   R_0(h;\ka)  &= \delta_{h,0}\\
  R_t(0;\ka) &= \ka R_{t-1}(1,\ka) \\
 R_t(h;\ka)&= R_{t-1}(h-1;\ka) + R_{t-1}(h+1;\ka).    
\end{align}\end{subequations}
The first and last equations are the same as for the unweighted Ballot paths and are satisfied by
$CT[\La^t z^h (1-z^2)g(z)]$ provided that on expansion $g(z)$ has no negative powers and $g(0)=1$
(which will be the case).
We have introduced the factor $g(z)$ to allow the second equation to be satisfied.  
But $z+1/z = \ka z + (1-(\ka-1) z^2)/z$ and choosing $g(z) = 1/(1-(\ka-1) z^2)$ 
\begin{eqnarray*}
CT[\La^t(1-z^2)g(z)]&=&
CT[\La^{t-1}(z+1/z)(1-z^2)g(z)]\\
 &=& \ka CT[\La^{t-1} z (1-z^2) g(z)] + CT[\La^{t-1}(1/z -z)].
\end{eqnarray*}
The result follows since the last term may be evaluated to give zero.
\end{proof}
\begin{cor}[\cite{BE}]
\begin{equation}
\label{eq:retpoly} R_{t}(h;\ka)=\sum_{m=0}^{\frac 12 (t-h)} B_{t,h+2m}(\ka-1)^m=
\sum_{m=0}^{\frac 12 (t-h)}B_{t-m-1,m+h-1}\ka^{m}.
\end{equation}
\end{cor}
\begin{proof}
The first equality follows by expanding \eqref{eq:rpctf} in powers of $(\ka-1)z^2$ and using \eqref{CTBallot}. The second is obtained by
rewriting the constant term formula as
\begin{equation}
R_t(h;\ka) = CT[\frac{\La^{t-1} z^{h-1} (1-z^2)}{1- \ka z/\La}]
\end{equation}
and then expanding in powers of $\ka z/\La$.
\begin{figure}[ht!]
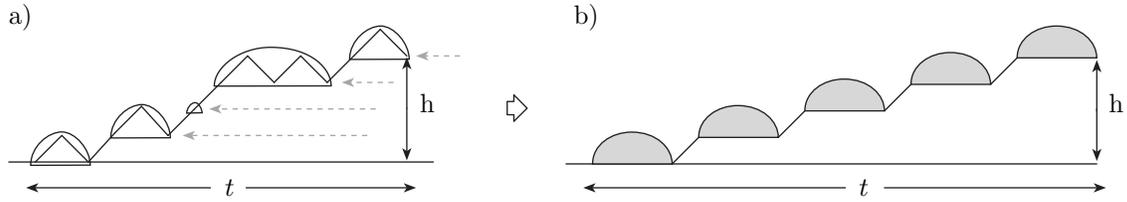

\centering
\ballot
\caption{\it  a)  An example of a Ballot path showing the ``terraces'' which
define the Bubbles shown schematically in b) which represents of a  Ballot
path of length $t$ and height $h=4$.  }
\label{fig:bal}
\end{figure}
This result suggests a bijection between Ballot paths of length $t$ with $m$ returns and 
Ballot paths of length $t-m-1$ and height $m+h-1$ and hence suggests a combinatorial proof.  
Such a bijection was given in \cite{BE2}.
 A Ballot path can be represented schematically as shown in figure
\ref{fig:bal}. The  Bubbles (see definition \ref{defn:bub}) represent a, possibly empty,  arbitrary elevated 
Dyck path. Using this schematic representation, proving \eqref{eq:retpoly} is then
straightforward. For simplicity we show the bijection in the case $h=0$ in figure \ref{fig:bp}.
\begin{figure}[ht!]
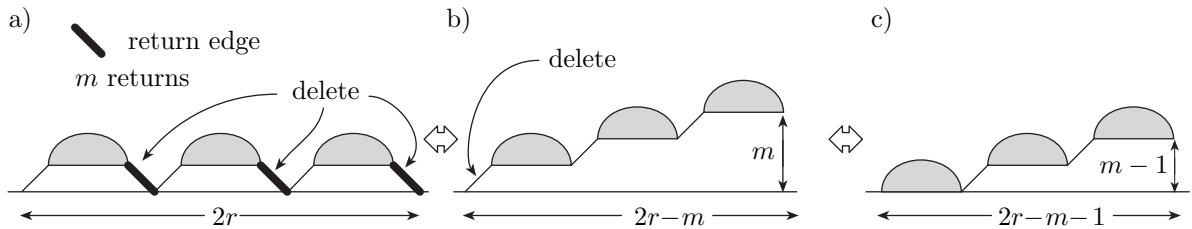

\centering
\returns
\caption{\it  a)  Schematic representation of a Dyck path with $m=3$ returns. b)  All   return steps deleted c) First step deleted produces a Ballot path of height $m-1$ and length $2r-m-1$.}
\label{fig:bp}

\end{figure}
The expansion in powers of $\ka-1$ was also obtained by bijection in \cite{BE2}. In this case the bijection
is between Ballot paths with at least $m$ returns, $m$ of which are marked, and Ballot paths of the same length
but of height $h+2m$.
\begin{figure}[ht!]
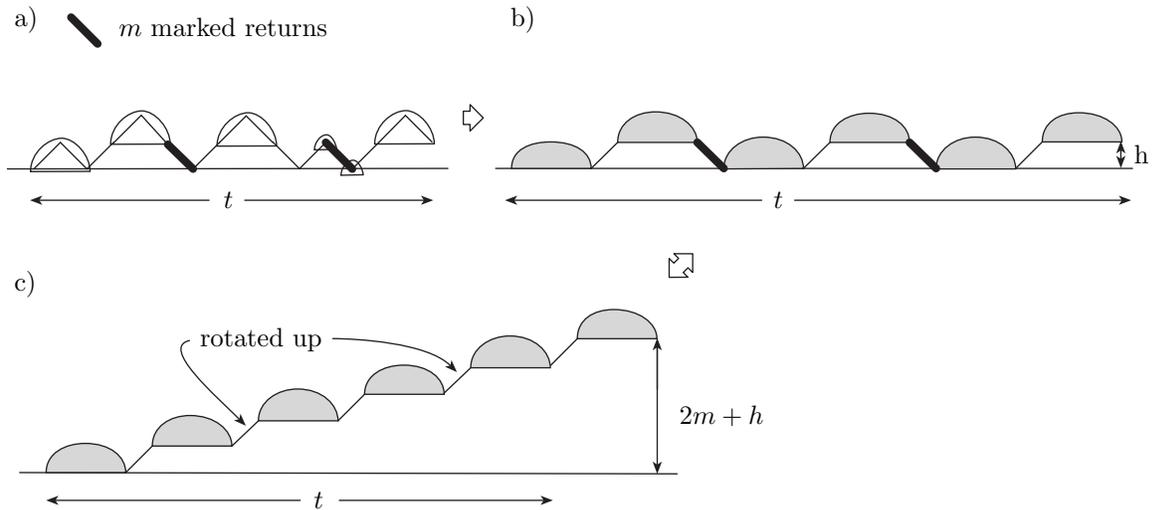

\centering
\markedBij
\caption{\it a) An example of a marked return Ballot path of height $h=1$ with $m=2$ marked returns and b) its schematic representation in terms of bubbles.  c)
Each marked return rotated through 90 degrees anti-clockwise (or equivalently replaced by an up step)    producing   a  Ballot path of
height $2m+h$ and length $t$.}
\label{fig:bpex}
\end{figure}
\end{proof}

The above methodology is typical of that used for the more complicated two parameter case in the 
next section. A constant term formula will be derived and then rewritten in four different ways
each giving rise to an expansion in pairs of different variables the coefficients of which are 
Ballot numbers and are
shown, by bijection, to enumerate various types of lattice path. 

\begin{figure}[ht!]
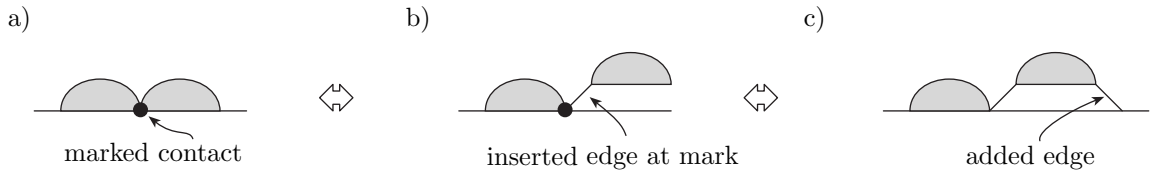

\centering
\markedContact
\caption{\it  a) Dyck (or hovering) path with one marked contact, b) insert an
up step at the mark produces a unique height one Ballot path (length $2r +1$)
and c) adding a final down step produces a unique Dyck path of length $2r+2$}
\label{fig:mc}
\end{figure}

In the next section we will also need the following lemma.
\begin{lemma} \label{lem:ret}
Let $\SepHovering_{2r;2s}$ be the set of Separated Hovering paths (definition \ref{def:mh} ) then
\begin{equation}
\label{markedhoversum}
\sum_{ s=0}^{r}|\SepHovering_{2r;2s}| = | \Dyck_{2r+2}|
\end{equation}
\end{lemma} 
\begin{proof} Since we can shift any Hovering path down on to the $x$-axis to give
a Dyck path with vertex $(2s,0)$ marked, the lemma says the number of Dyck paths with one contact
marked is equal to the number of (unmarked) Dyck paths two steps longer. A
sketch of the bijective combinatorial proof is shown in figure \ref{fig:mc}.
The construction is a bijection, since, given any length $2r+2$  Dyck path a 
unique 
length $2r $ marked contact Dyck path is determined by deleting the rightmost
 step and the rightmost step from $y=0$ to $y=1$ (and marking the left vertex of
the latter step).
\end{proof}

\section{The two contact model.}

We begin by computing  $Z_{2r}^{(3)}$ using the $W^{(4)}$ weights -- we will
refer to this as the two contact model. We will generalise the model by
allowing arbitrary starting and ending heights for the paths. In particular, let
the paths be of length $t$, start  at $(0,y^i)$ with $y^i \in \integer_{odd}$
and terminate  at $(t,y^f)$, $t+y^f \in \integer_{odd}$, i.e. Cross paths in $P^{(C)}_{t;y^i,y^f}$ .

This model can also be thought of as a polymer model with two different surface
interactions, or ``contacts''. The path interacts with the ``thickened'' surface
$y=0,1$ via two parameters, $\k_1$ and $\k_2$ and the partition function is
defined by
\begin{equation}
\pf_t(y^{f}|y^i;\k_1,\k_2) = \sum_{\om \in P^{(C)}_{t;y^i,y^f}} W^{(4)}(\om)
\end{equation}
where the sum is over Cross paths of length $t$ with given initial and final
heights $y^i$ and $y^f$ and the weight $W^{(4)}(\om)$ is defined by
\eqref{eq:449} and \eqref{w4}. This is a generalisation of the
ASEP partition function $Z_{2r}^{(3)}$ thus
\begin{equation}
Z_{2r}^{(3)}=\pf_{2r}(1|1;\ab\bb,\k^2)
\end{equation}

This generalisation allows this partition function to be determined by recurrence relations similar to those
for the return polynomial of Ballot paths. 
By considering paths of length $t-1$ which can reach the point $(t,y)$ by adding one more step
the   partition function $\pf_t(y^{f}|y^i;\k_1,\k_2)$ may be seen to satisfy the equations,
\begin{align}
\pf_1(1|y^i;\k_1,\k_2) &=0\label{Zdef1}\\
\pf_0(y|y^i;\k_1,\k_2) &= \delta_{y,y^i}\label{Zdef2}\\
\pf_t(0|y^i;\k_1,\k_2) &= \pf_{t-1} 
(1|y^i;\k_1,\k_2)\label{bc1}\\
\pf_t(1|y^i;\k_1,\k_2) &= \k_1\pf_{t-1} 
(0|y^i;\k_1,\k_2)+\k_2\pf_{t-1} (2|y^i;\k_1,\k_2),\qquad t\geq 
2\label{bc2}\\
\intertext{and for $t=1,2,\dots$; $y=2,3,\dots$ and $t+y$ odd}
\pf_t(y|y^i;\k_1,\k_2) &= \pf_{t-1}(y-1|y^i;\k_1,\k_2) + 
\pf_{t-1}(y+1|y^i;\k_1,\k_2).\label{general}
\end{align}

These partial difference equations can be solved for
$\pf_t(y^{f}|y^i;\k_1,\k_2)$ by  using the constant term method.

\begin{prop}\label{prop1} With $\bar \k_i =\k_i-1$, $\bar z = 1/z$ and $\Lambda = z + \bar z$, for $y^i \in
\integer_{odd}$ and $y\ge 1$
\begin{equation}
\label{CTZ}
\pf_t(y|y^i;\k_1,\k_2) = CT[\La^t(z^{y-y^i}-z^{y+y^i-2})]+  
(\delta_{y^i,1}+\k_2(1-\delta_{y^i,1}))CT[z^{y+y^i-2}\La^t G(z)]
\end{equation}
where
\begin{equation}\label{Gz}
G(z) = \frac {(1-z^2)z \La}{1 - (\bar \k_1 + \bar \k_2)z^2 - \bar \k_2  z^4}
\end{equation}
{\it Note:}\,
For $y^i=0$ and $y\ge 1$
\begin{equation}
\pf_t(y|0;\k_1,\k_2) = \k_1 \pf_{t-1}(y|1;\k_1,\k_2)
\end{equation}

\end{prop}
\begin{proof} Substituting \eqref{CTZ} into the partial difference equations and
noting that $G(z)$ may be expanded in even powers of $z$ with no inverse powers shows
that \eqref{Zdef1}, \eqref{Zdef2} and \eqref{general} are satisfied. \eqref{bc1}
may be taken as the definition of $\pf_t(0|y^i;\k_1,\k_2)$ and \eqref{bc2} may
then be transformed into
\begin{equation}\label{modZ}
\pf_t(1|y^i;\k_1,\k_2) = \k_1\pf_{t-2} 
(1|y^i;\k_1,\k_2)+\k_2\pf_{t-1} (2|y^i;\k_1,\k_2),\qquad t\geq 2
\end{equation}

To verify that this equation is also satisfied we note that
\begin{equation}
\La^2 G(z) = \La(1/z-z)+ (\k_1 + \k_2 z \La )G(z)\label{La2}
\end{equation}

In the case $y^i =1$ this gives
\begin{align*}
\pf_t(1|1;\k_1,\k_2) &= CT[\La^t G(z)]
&= CT[\La^{t-1}(1/z-z)]
+\k_1\pf_{t-2} (1|y^i;\k_1,\k_2)+\k_2\pf_{t-1} (2|y^i;\k_1,\k_2)
\end{align*}
and \eqref{modZ} follows since the first term is zero.

Otherwise $y^i=3,5,\dots$ in which case
$$
\pf_{t-1}(2|y^i;\k_1,\k_2) = CT[(z^{2-y^i}- z^{y^i})\La^{t-1}] + \k_2 CT[z^{y^i}\La^{t-1} G(z)]
$$
and using this together with \eqref{La2} and the fact that the first term in \eqref{CTZ} vanishes
when $y=1$
\begin{align*}
\pf_{t}(1|y^i;\k_1,\k_2) &= \k_2 CT[z^{y^i-1} \La^t G(z)]\\
&=\k_2 CT[z^{y^i-1}(1/z-z)\La^{t-1}]+\k_1\pf_{t-2} (1|y^i;\k_1,\k_2) +\k_2^2CT[z^{y^i}\La^{t-1}G(z)]\\
&=\k_2 CT[(z^{y^i-2}-(1/z)^{y^i-2})\La^{t-1}]+\k_1\pf_{t-2} (1|y^i;\k_1,\k_2) +\k_2\pf_{t-1} (2|y^i;\k_1,\k_2)
\end{align*}
and again \eqref{modZ} follows since the first term evaluates to zero.

An alternative constructive proof of this proposition is given in appendix B.
\end{proof}

\begin{cor} \label{crosspaths}
The number of $t-$step Cross paths which begin at $v^i = (0,h_1)$
and end at $v^f=(t,h_2)$ 
is given by
\begin{align}
\label{kappa1}
\pf_t(h_2|h_1;1,1) &= \binom{t} {\frac{1}{2}
(t-h_2+h_1)}-\binom{t}{\half(t-h_2-h_1)-1}
\intertext{in terms of which we can write, $\pf_t(y|y^i;\k_1,\k_2)$ as  }
\pf_t(y|y^i;\k_1,\k_2) &=
\pf_t(y-2|y^i-2;1,1)+(\delta_{y^i,1}+\k_2(1-\delta_{y^i,1}))\Za_t(y|y^i;\k_1,\k_2)\label{anchored}\\
\intertext{where }
\label{Za}
\Za_t(y|y^i;\k_1,\k_2)&= CT\left[\La^t z^{y+y^i-2} G(z)\right]\\
\intertext{is the partition function restricted
to Anchored Cross paths except that for $y^i>1$ the step leading to the first visit to $y=1$ has weight $1$.}
\intertext{In particular}
\label{Z11}
\pf_t(y|1;\k_1,\k_2) &= \Za_t(y|1;\k_1,\k_2) =CT\left[\La^t G(z)z^{y-1}\right].
\end{align}

\end{cor}
\noindent{\it Notes:} 
\begin{itemize}
\item
It follows from the constant term formula that
\begin{equation}
\Za_t(y|y^i;\k_1,\k_2)  = \pf_t(y+y^i-1|1;\k_1,\k_2)
\end{equation}

\item
 Setting $\k_1=\k_2=1$ gives the number of unweighted paths starting at height $1$.
$$\pf_t(y|1;1,1) = CT[\La^t(1-z^4)z^{y-1})=CT[\La^{t+1}(1-z^2)z^y] = B_{t+1,y}$$
as expected since the paths biject to Ballot paths by adding an initial up step.
\end{itemize}

\begin{proof} With $\k_1=\k_2 =1$, $G(z) = 1-z^4$ and substituting in
\eqref{CTZ} gives
\begin{equation}\label{Zone}
\pf_t(y|y^i;1,1) = CT\left[\La^t (z^{y-y^i} -z^{y+y^i+2})\right]
\end{equation}
which yields \eqref{kappa1}.
 
$\pf_t(y-2|y^i-2;1,1)$ is the number of Cross paths which avoid $y=1$. 
This follows since such paths are in simple bijection with 
the paths starting at height $y^i-2$ and ending at height $y-2$, eg.\  
just push the whole path down (or up)  two units. The second term in 
\eqref{anchored} is therefore the partition function for Anchored Cross paths.
In the case $y^i >1$ Anchored Cross paths always have a first visit to $y=1$
and the step leading to this visit has a factor $\k_2$. Removing this factor
leaves $\Za_t(y|y^i;\k_1,\k_2)$ which is therefore the partition function for
Anchored Cross paths except that for $y^i>1$ the step leading to the first visit 
to $y=1$ has weight $1$. Setting $y^i=1$ in \eqref{anchored} gives \eqref{Z11} since the first term vanishes.

\end{proof}

\begin{cor}\label{cor:bp3}
\begin{align}\label{eq:sdf}
\Za_t(y|y^i;\k_1,\k_2)&=CT\left[\frac{(1-z^2)\Lambda^{t-1} z^{y+y^i-3}}
{1-(\k_1+z\Lambda \k_2)(z/\Lambda)^2}\right]\\
&\notag\\
&=\sum_{j=0}^{(t-y-y^i)/2+1}\quad\sum_{k=0}^{(t-y-y^i)/2-j+1}\k_1^j \k_2^k\,
\binom{j+k}{k}\, B_{t-2j-k-1,y+y^i+k-3}\label{k12}
\end{align}
\end{cor} 
{\it Note:}\,When $y=y^i=1$ and $j = t/2, k=0$ the coefficient
$B_{-1,-1}$ is indeterminate but setting $B_{-1,-1}=1$ gives the correct answer.
This corresponds to the path which alternates between $y=0$ and $y=1$. Notice
that replacing the factorials in the definition of $B_{t,h}$ by Gamma functions
gives 
$B_{t,t}=1$ for all $t>-1$.

\noindent
\begin{proof} Rewrite $G(z)$ in the form $$
G(z) = \frac{1-z^2} {(z\Lambda)(1-(\k_1+z\Lambda \k_2)(z/\Lambda)^2)},$$
expand in powers of $\k_1$ and $\k_2$ and use the CT formula \eqref{CTBallot} for Ballot numbers.

Equation \eqref{k12} may also be proved by a combinatorial argument.
 To simplify the proof we only consider the ASEP case $y=y^{i}=1$. The extension to general $y$,
$y^i$ is straightforward. Recall that,
from the weight definition, \eqref{w4}, the returns to $y=1$ are weighted with
$\ka_1$ or $\ka_2$ depending on whether the return is from below or from above $y=1$.
The binomial
coefficient corresponds to
choosing a particular sequence of
$\ka_{1}$ and $\ka_{2}$ weighted returns. For each particular sequence of returns we need
to show there are $B_{t-2j-k-1,k-1}$ possible path configurations. 

We first represent 
a particular sequence schematically and
then show any path corresponding to the schematic can be bijected to a Ballot
path with the correct height and length. Schematically an example of a
particular sequence of $\ka_{1}$ and $\ka_{2}$ returns is shown in figure
\ref{fig:k1k2seq}.
\begin{figure}[ht!]
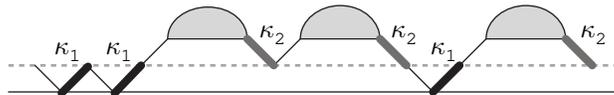

\centering
\kkseq
\caption{\it  Schematic representation of a one-up path corresponding to the
return sequence $\ka_1^{2}\ka_2^{2}\ka_1\ka_2$. }
\label{fig:k1k2seq}
\end{figure}

We now perform three operations to biject a given sequence into a Ballot path.
\begin{itemize}
\item  First delete all $\ka_{2}$ return steps, see figure \ref{fig:k1k2bij}a).
This produces a path of length $t-k$ and height $k+1$.
\item Next, delete the first up step
above $y=1$ (if any -- which is the case if $k>0$), see figure
\ref{fig:k1k2bij}b). This produces a path of length $t-k-1$ and height $k$.
\item Finally, delete all $2j$
steps originally below $y=1$, see figure \ref{fig:k1k2bij}c). This produces a
Ballot path of length $t-k-1-2j$ and height $k-1$ as required.
\end{itemize}
Given the sequence of $\ka_{1}$ and $\ka_{2}$ the reverse direction for the bijection is obtained by simply reversing the forward mapping -- see figure \ref{fig:k1k2bij}.
\end{proof}
\begin{figure}[ht!]
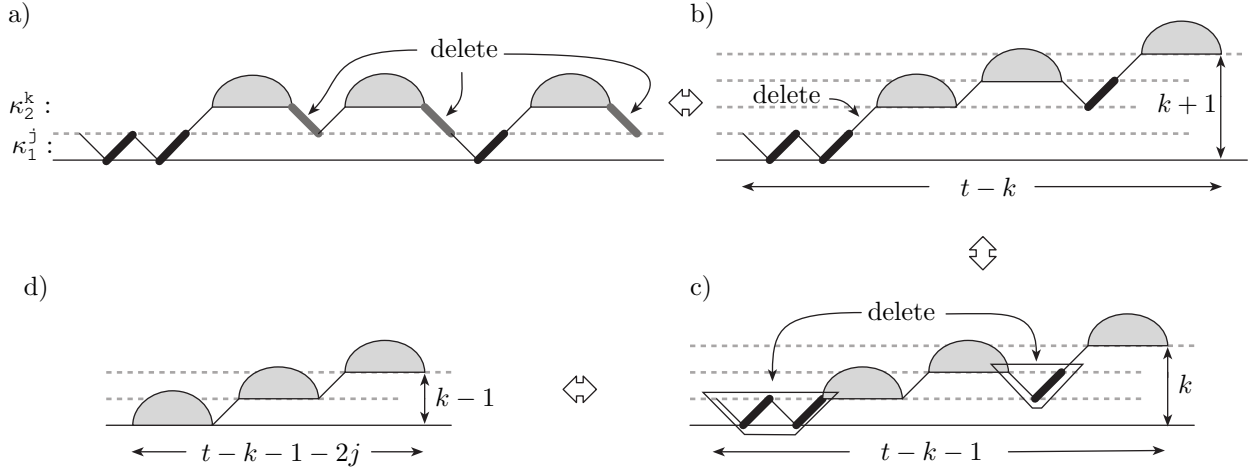

\centering
\kkbij
\caption{\it Bijection of first example sequence in figure \ref{fig:k1k2seq} to
a Ballot path: a) Delete all $\ka_{2}$ return steps, then b)  
delete first up step above $y=1$, then c)   delete all steps originally
below $y=1$, to give d) the final Ballot path. }
\label{fig:k1k2bij}
\end{figure}

\begin{cor}\label{cor:bp2}
\begin{equation}
\Za_t(y|y^i;\k_1,\k_2) 
=\sum_{j=0}^{(t-y-y^i)/2+1}\quad\sum_{k=0}^{(t-y-y^i)/2-j+1}\bar \k_1^j
\bar \k_2^k\, \binom{j+k}{ k}\, B_{t+k+1,y+y^i+2j+3k-1}\label{Z19}
\end{equation}
\end{cor}

\noindent
\begin{proof} 
Rewrite $G(z)$ in the form $$
G(z) = \frac{(1-z^2)z \Lambda}{1-(\bar\k_1+z\Lambda\bar\k_2)z^2}, $$
expand in powers of $\bar\k_1$ and $\bar\k_2$ and use the CT formula \eqref{CTBallot} for Ballot numbers.

As with corollary \ref{cor:bp3}, the result, \eqref{Z19},  may also be proved combinatorially as follows. Again, in
order to simplify the proof we consider only the ASEP case $y=y^i=1$. The substitution $\ka_i = 1+\bar\k_i$
means that a return which was weighted with $\ka_i$ is now weighted with either $\bar\ka_i$ or $1$. The factor
\begin{equation}
\label{eq:kbfac}
\binom{j+k}{k}\, B_{t+k+1, 2j+3k+1}
\end{equation}
is the number of paths with a subset of exactly $j$ of the
returns from below marked and exactly $k$ of the
returns from above marked corresponding to the choice of weight $\bar\ka_i$ or $1$.  
The binomial coefficient in  \eqref{eq:kbfac}
is the number of ways of choosing a particular sequence of
$\bar\ka_1$ and $\bar\ka_2$ weighted marks (reading the path from left to right), whilst,
for a given sequence, the Ballot number represents the number of paths
corresponding to the sequence.
\begin{figure}[ht!]
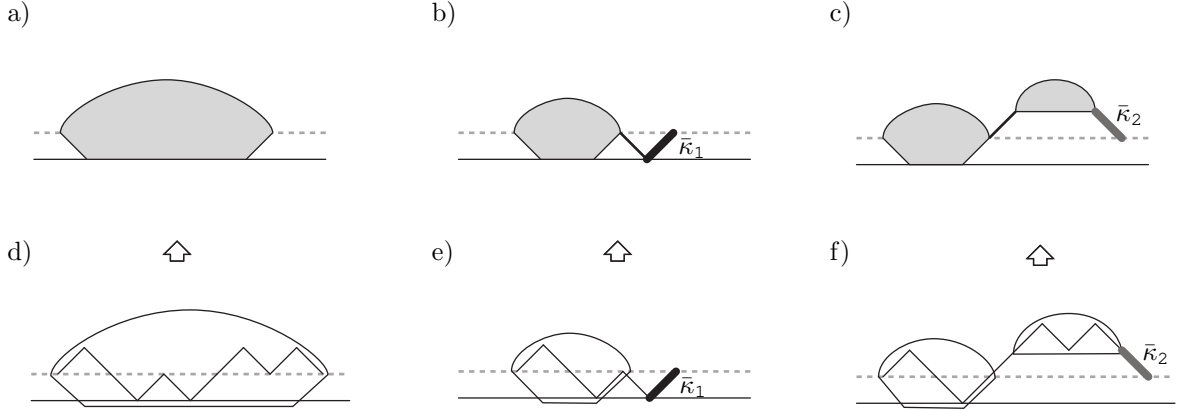

\centering
\fryingpans 
\caption{\it a) Schematic One Up path or ``frying pan'' with no steps marked. b) Schematic One Up path with only the last step $\bar\ka_1$ marked. c) Schematic One Up path with only the last step $\bar\ka_2$ marked. d ) An example showing how the schematic  frying pan  represents a, possibly empty, One Up path. e) An example showing a  frying
pan followed by a down step and an up step (which forms a $\bar\ka_1$ marked return).
f) An example showing a  frying
pan followed by an up step, then a Bubble then a final down step
(which forms a $\bar\ka_2$ marked return).  }
\label{fig:fps}
\end{figure}
 The most general path corresponding to a given sequence can be represented
schematically by concatenating the corresponding schematic sub-paths shown in
figure \ref{fig:fps}a) and
\ref{fig:fps}b) with a final ``frying pan'' shown in figure \ref{fig:fps}c). 
Examples of sub-paths corresponding to the three types of schematics are
illustrated in figures \ref{fig:fps}d) -- \ref{fig:fps}f). Note, the shaded
regions of the schematics represent any number (possibly zero) of steps.
 Examples of two possible sequences are illustrated in  figure \ref{fig:kbs}.

Thus, for a given return sequence we need to show that there are  
$B_{t+k+1, 2j+3k+1}$  return marked paths. Without loss of generality
we choose a typical sequence and represent it schematically as shown in figure
\ref{fig:kbs}. 
\begin{figure}[ht!]
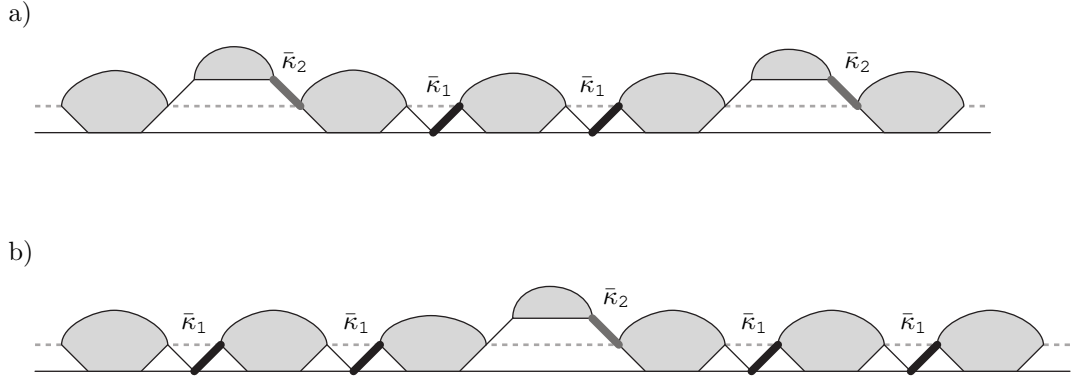

\centering
 \kbseq
\caption{\it  Schematic $\bar\ka_1$ -- $\bar\ka_2$ marked hovering path. a) For
the sequence $ \bar\ka_2  \bar\ka_1  \bar\ka_1  \bar\ka_2 $  and b) for the
sequence $ \bar\ka_1  \bar\ka_1  \bar\ka_2  \bar\ka_1 \bar \ka_1$.}
\label{fig:kbs}
\end{figure}

Thus to prove the Ballot number factor in \eqref{eq:kbfac} we need to biject any
schematic marked return sequence  to a Ballot path
of length $t+k+1$ and height $2j+3k+1$. We do this by bijecting each schematic in
the sequence   to a sub-Ballot path (plus, possibly, an extra step) and then
concatenate them all together.
\begin{itemize}
\item  Thus,
the last frying pan, of length say, $2r'$, bijects to a  height $1$, length $2r' + 1$
Ballot path -- see figure \ref{fig:dfg}.
\begin{figure}[ht!]
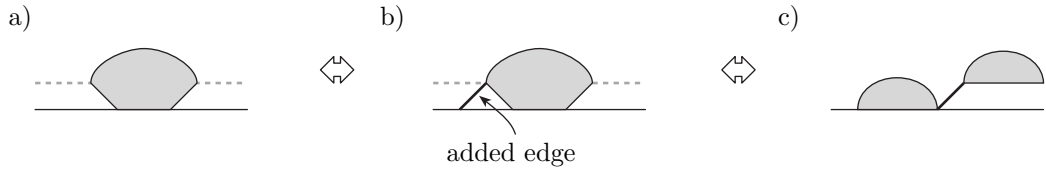

\centering
 \lastfp
\caption{\it  Bijection of a ``frying pan'' to a height one Ballot path. The
final path is one step longer than the original.}
\label{fig:dfg}
\end{figure}
\item A marked $\bar\ka_2$ schematic of length $2r_1$ bijects to a height $2$,
Ballot path with an additional final up step (see figure
\ref{fig:kbo}). The final length is  
$2r_1+1$ as an extra step has  to be added.
\begin{figure}[ht!]
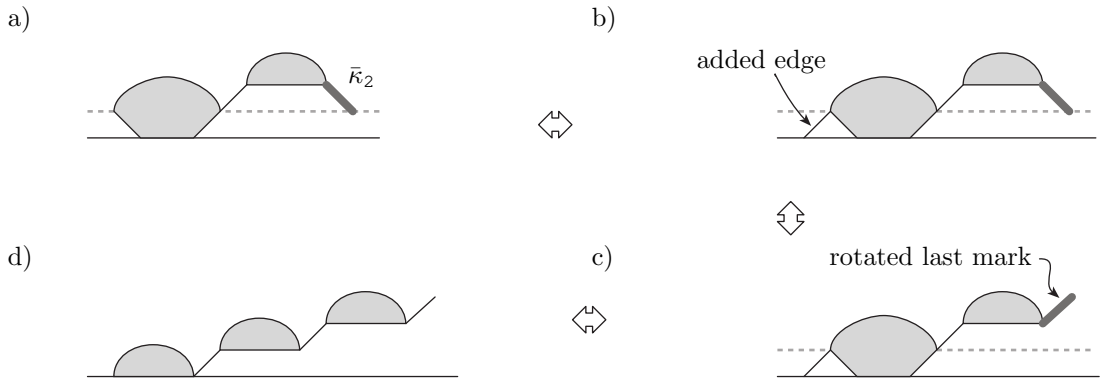

\centering
 \kbo
\caption{\it Bijection of a $\bar\ka_2$ schematic  to a height two Ballot path,
with a final up step. The final path is one step longer than the original.}
\label{fig:kbo}
\end{figure}
\item Finally, a marked $\bar\ka_1$ schematic of length $2r_2$ bijects to a height
$1$, Ballot path with an additional final up step (see figure
\ref{fig:kbt}). The final length is unchanged.
\begin{figure}[ht!]
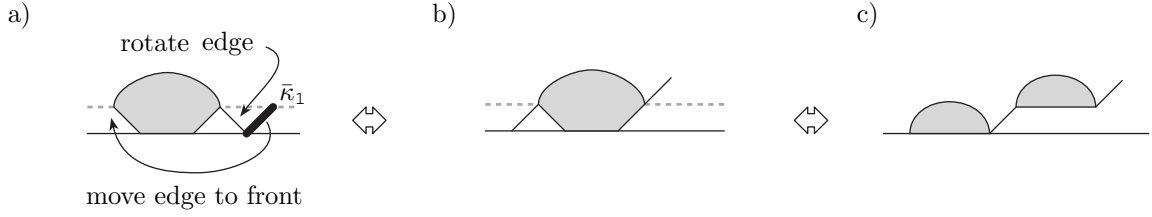

\centering
 \kbt
\caption{\it  Bijection of a $\bar\ka_1$ schematic  to a height one Ballot path,
with a final up step.  }
\label{fig:kbt}
\end{figure}
\end{itemize}
Putting these moves all together is illustrated in figure \ref{fig:kbokbt},
which show clearly a Ballot path of length $t+k+1$ and height $2j+2k+k+1$ is
obtained.
\begin{figure}[ht!]
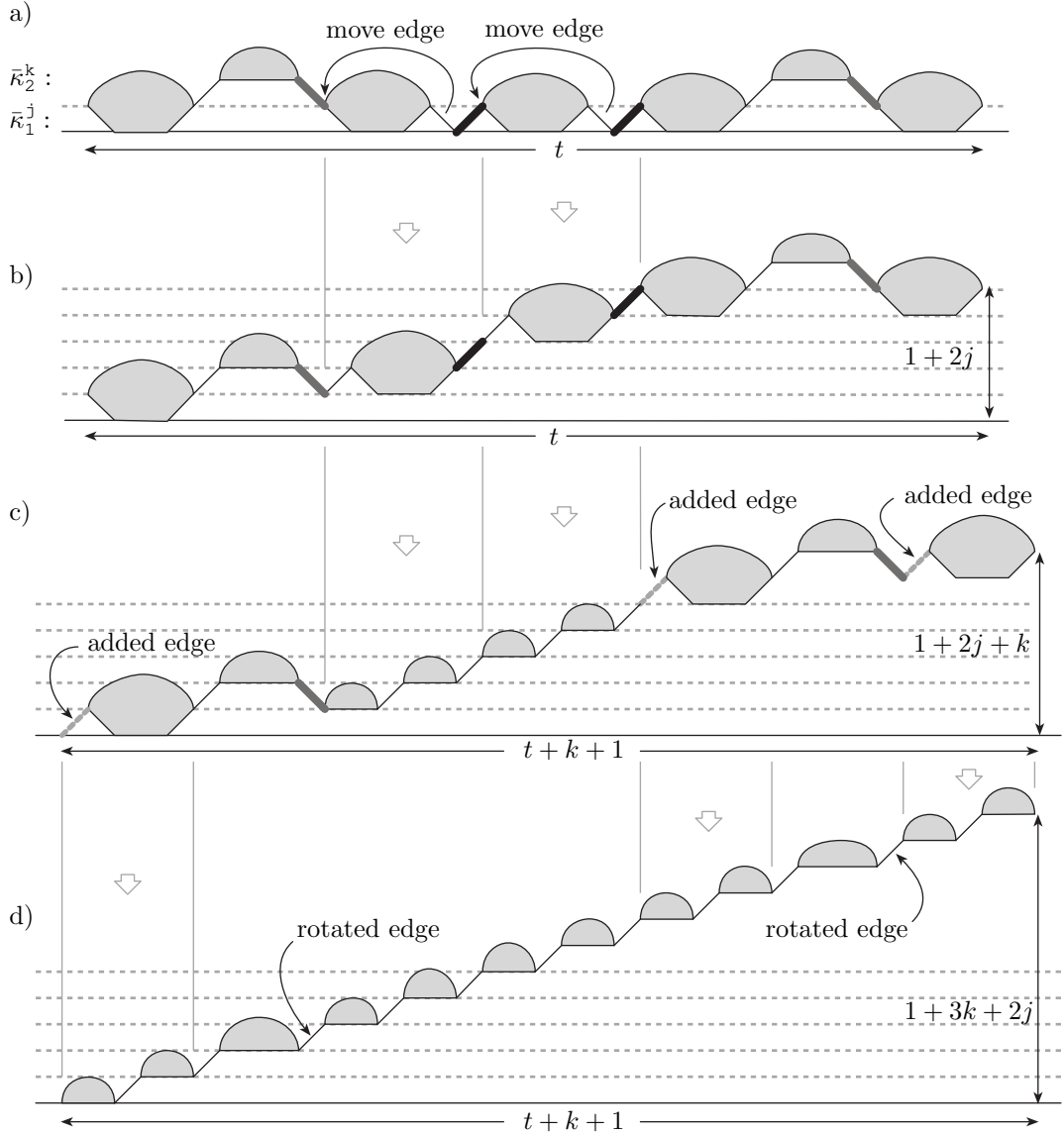

\centering
\kbokbt
\caption{\it  Bijection between a One Up path with a given sequence of marked $\bar\ka_{1}$ and $\bar\ka_{2}$ steps and a Ballot path. a) The marked One Up path. b) After the application of move  illustrated in figure \ref{fig:kbt}. c)  After the application of move  illustrated in figure \ref{fig:kbo}. e) Final Ballot path after rotating the remaining marks. }
\label{fig:kbokbt}
\end{figure}

\end{proof}

An explicit evaluation of the partition function of the canonical path representation, defined by \eqref{eq:4422},
 is the
case $y^i=y^f=1$ of the following result.
\begin{cor}\label{alphabeta} With $\k_1= \ab \bb$ and $\k_2 =\ab+\bb-\ab \bb$
\begin{align}
\Za_{t}(y^i|y^f;\k_1,\k_2)&= \sum_{m=0}^{\frac 12 (t-y^i-y^f)+1} B_{t-m-1,m+y^i+y^f-3}\,\,\,\frac{\bar
\alpha^{m+1}-\bar \beta^{m+1}}{\bar\alpha-\bar\beta}\\
&= \sum_{m=0}^{\frac 12 (t-y^i-y^f)+1} B_{t-m-1,m+y^i+y^f-3}\,\,\,\sum_{j=0}^m\bar \alpha^j \bar
\beta^{m-j}\label{Zabbb}
\end{align}
\end{cor}
\noindent{\it Notes:}\, 
\begin{itemize}
\item
For the ASEP model ($y^i=y^f=1, t=2r$) with $r\ge1$, \eqref{Zabbb} reduces to the result of 
Derrida et al \cite{DEHP} equation (39)
\begin{equation}
\pf_{2r}(1|1;\k_1,\k_2)= \sum_{m=1}^r
\frac{m(2r-m-1)!}{r!(r-m)!}\sum_{j=0}^m\bar \alpha^j \bar \beta^{m-j} = Z_{2r}^{(5)}
\end{equation}
which they obtained directly from the algebra \eqref{algebra} (\cite{DEHP} appendix A1,
equation (A12)) without the use of a matrix representation. 
\item
The result for the ASEP model may be written in terms of the return polynomial for Ballot paths, thus
\begin{eqnarray}
\pf_{2r}(1|1;\k_1,\k_2)&=& \sum_{j=0}^r \bar \a^j \sum_{\ell=0}^{r-j}B_{2r-j-\ell-1}{j+\ell-1}\bar \b^\ell\nonumber\\
&=& \sum_{j=0}^r R_{2r-j}(j;\bar\b) \bar\a^j.
\end{eqnarray}
This formula may also be derived using a path representation based on recurrence relations of Derrida,
Domany and Mukamel \cite{DDM}.
\end{itemize}

\begin{proof}
By definition of $\ab$ and $\bb$
\begin{equation}\label{Gden}
1 - (\bar \k_1 + \bar \k_2)z^2 - \bar \k_2  z^4 = [1-(\ab-1)z^2][1-(\bb-1)z^2]=z^2(\La-\ab z)(\La -\bb z)
\end{equation}
and from \eqref{Gz}
\begin{equation}\label{Gabbb}
G(z) = \frac{(1-z^2)\bar z \La}{(\La-\ab z)(\La -\bb z)}= \frac{(1-z^2)\bar z^2}{\ab-\bb}\left(\frac 1{1-\ab z/\La}-
\frac 1{1-\bb z/\La}\right)
\end{equation}
The result follows by expanding in powers of $\ab$ and $\bb$,
substituting in \eqref{Za} and using the CT formula \eqref{CTBallot} for Ballot numbers.

In the ASEP case $y^i=y^f=1, t=2r$ the coefficient $B_{2r-m-1,m-1}$ in \eqref{Zabbb} is equal to the number 
of Dyck paths with $m$ returns to $y=0$ (see \eqref{eq:retpoly}).  The equality
of $\pf_{2r}(1|1;\k_1,\k_2)$ with $Z_{2r}^{(5)}$, defined by \eqref{eq:4422}, follows by raising 
the Dyck paths so that they become
hovering paths $\om \in P^{(H)}_{2r}$ with returns to $y=1$. 
The factor $\ab^j \bb^{m-j}$  in \eqref{Zabbb} corresponds
to weighting the first $j$ returns of $\om$ with $\ab$ and the remainder with $\bb$.
The sum over $s$ in \eqref{eq:4422} is obtained by partitioning the weighted paths according to the position $(2s,1)$
of the $j^{th}$ return (ie.\ the separation vertex).
 
The equality of $Z_{2r}^{(3)}$ and $Z_{2r}^{(5)}$ is also shown directly in \cite{BER} by involution.
\end{proof}

\begin{cor} Let $\ab = c+1$ and $\bb = d+1$ then $\k_1=(c+1)(d+1)$, $\k_2 =
1-cd$ and
\begin{align}
\Za_{t}(y|y^i;\k_1,\k_2) &= \sum_{m=0}^{(t-y-y^i+2)/2} 
B_{t+1,y+y^i+2m-1}\sum_{j=0}^m c^j d^{m-j}\label{Zac}\\
&= \sum_{m=0}^{(t-y-y^i+2)/2} B_{t+1,y+y^i+2m-1}\frac
{(c^{m+1}-d^{m+1})}{c-d}\\
\intertext{in particular, in the ASEP case}
Z_{2r}(1|1;(c+1)(d+1),1-cd) &= Z_{2r}^{(2)} = \sum_{m=0}^r 
B_{2r+1,2m+1}\sum_{j=0}^m c^j d^{m-j} \label{Z11cd}
\end{align}
\end{cor}

{\it Note:} 
Equation \eqref{Z11cd} is not given in \cite{DEHP} but (34) 
and (35) of \cite{DEHP} together give the related formula
\begin{equation}\label{cdfactor}
<W|C^r|V>=(1- c d)\sum_{i=1}^\infty\sum_{j=1}^\infty 
\left(\binom{2r}{r+i-j} - \binom{2r}{r+i+j}\right)c^{i-1} d^{j-1}
\end{equation}
This expression involves an infinite series whereas our expression is finite.
By corollary \ref{crosspaths} the coefficient in \eqref{cdfactor}
is the number of Cross paths of length $2r$ with 
$h_1= 2i-1$ and $h_2=2j-1$ and the double sum extends over $P_{2r}^{(C)}$.
The factor $1-cd$ restricts the sum to Anchored Cross paths.
The equivalence of \eqref{cdfactor} and \eqref{Z11cd} is shown in \cite{BER}
by constructing an involution on $P_{2r}^{(C)}$ having $P_{2r}^{(aC))}$
as is its fixed point set.

\begin{proof}
 From \eqref{Gden}
$$1 - (\bar \k_1 + \bar \k_2)z^2 - \bar \k_2  z^4 = (1-c z^2)(1-d z^2)$$
and substitution in \eqref{Gz} gives 
\begin{equation}
\label{Gcd} G(z) = \frac {(1-z^2)z \La}{(1-c z^2)(1-d z^2)}
\end{equation}
The result follows
from \eqref{Za} using the expansion
\begin{equation}
\label{Gcdex}
\frac1{(1-cz^2)(1-dz^2)} = \frac 1{c-d}\left(\frac c{1-cz^2} -\frac
d{1-dz^2}\right),
\end{equation}
expanding in powers of $z$ and using the CT formula \eqref{CTBallot} for Ballot numbers..

Again a  combinatorial proof is possible which for simplicity we only give in the ASEP case.
The equality with $Z_{2r}^{(2)}$ follows from corollary \ref{Z2Z5}. To obtain the Ballot
number formula we obtain a bijection between $\bigcup_{s = 0}^r P^{(mH)}_{2r;2s}$ and the
set of Ballot paths of length $2r+1$ and height $h=2m+1$.

 An example of the schematic representation of a path $\om_s \in P^{(mH)}_{2r;2s}$ with a
given number of $c$ and $d$
marked return steps and a separating vertex $v_{2s}$ is shown in figure \ref{fig:cdseq}.  
As there are no steps below $y=1$ we have pushed the
hovering path down by unit height and consider it as a Dyck path.
\begin{figure}[ht!]
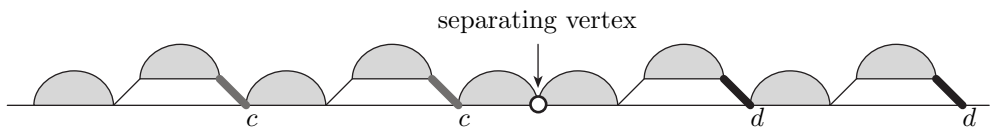

\centering
\cdseq
\caption{\it  An example of a schematic representation of a Dyck path with a
given number of $c$ and $d$ marked returns and a  marked separating vertex $v_{2s}$. }
\label{fig:cdseq}
\end{figure}

As in the proof of lemma \ref{lem:ret}, inserting an up step in each path at the 
position of $v_{2s}$
and taking the union over all possible positions of $v_{2s}$ starting at the last $c$ return and up to
but not including the first $d$ return (or from the beginning if there are no
$c$ returns and to the end if there are no $d$ returns) replaces the  pair of Bubbles on
either side of the separating vertex by the set Ballot paths of height one and one
step longer, see figure \ref{fig: cdbij}a,b). Replacing each marked return step
with an up step (and hence increasing the height  of the path by 2 each time)
then produces a Ballot path of length $2r+1$ and height $2m+1$ as required, see figure
\ref{fig: cdbij}c).
\begin{figure}[ht!]
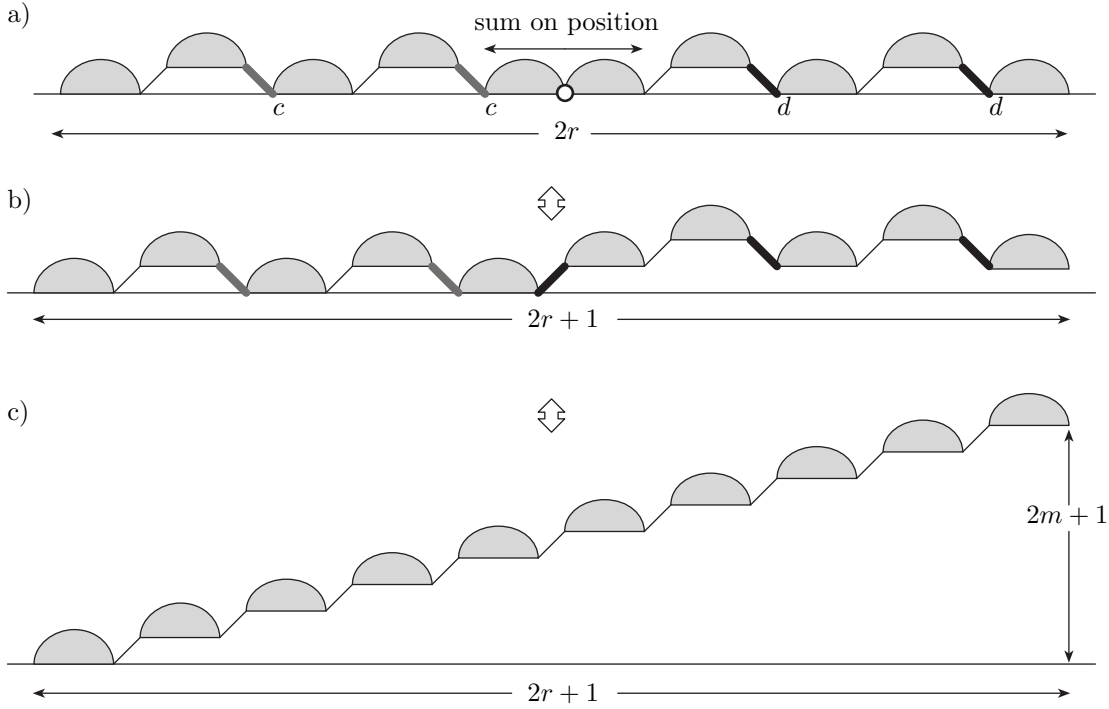

\centering
\cdbij
\caption{\it  a) Taking the union over positions of the marked separating vertex 
produces a schematic height one
Ballot path as in b), replacing each marked $c$ and $d$ return by an up step
produces a Ballot path of length $2r+1$ and height $2m+1$ as in c).  }
\label{fig: cdbij}
\end{figure}

\end{proof}

\begin{cor} \label{cor6}
$$\pf_{2r}(1|1;\k_1,\k_2)= -\frac 12(1-cd)CT\left[ \frac {\Lambda^{2r}(z^2-\bar
z^2)^2}{((1+c)^2 -c
\Lambda^2)((1+d)^2-d\Lambda^2)}\right] $$
\end{cor}
\noindent{Note:}\,This converts to the integral formula of Derrida et al \cite{DEHP} (B10) with $z^2 =
e^{i\theta}$ and and using a contour integral to pick out the constant term. 
It is related to the $\om$ expansion (see later). (B10) was obtained by finding
the eigenvectors of $C_2$ 
and is therefore an evaluation of $Z^{(2)}$.
\begin{proof} With $w\equiv z^2$ in \eqref{Gcd} and using \eqref{Z11} $$
\pf_{2r}(1|1;\k_1,\k_2) = CT\left[\frac
{\Lambda^{2r}(1-w^2)}{(1-cw)(1-dw)}\right] $$
Now symmetrize the denominator by multiplying numerator and denominator by $(1-c
\bar w)(1-d \bar w)$.

The result follows using $$
(1-w^2)(1-c \bar w)(1-d \bar w)= 1-cd -w^2 +cd \bar w^2 +(c+d)(w-\bar w) $$
Because the rest of the expression is now symmetric the contribution from the 
last term vanishes by replacing $\bar w$ 
by $w$ and both $w^2$ and $\bar w^2$ can be replaced by $(w^2+\bar w^2)/2$.
\end{proof}

\section{The $``\omega"$ expansion and phase diagram of the ASEP model.}

With $\omega_c=c/(1+c)^2$ corollary \ref{cor6} may be written in the form
\begin{equation}\label{Zomcd}
\pf_{2r}(1|1;\k_1,\k_2)=\frac{\oc -\od}{c-d}
CT\left[\frac{\Lambda^{2r+2}(1-z^2)}{(1-\oc \Lambda^2)(1-\od \Lambda^2)}\right].
\end{equation}
which may be expanded to give
\begin{equation}
\pf_{2r}(1|1;\k_1,\k_2)= \frac{\oc Z_{2r+2}(\oc)-\od Z_{2r+2}(\od)}{c-d}
\end{equation}
where
\begin{equation}\label{Zom}
Z_{2r}(\om) = CT\left[\frac{\La^{2r}(1-z^2)}{1-\om \La^2} \right].
\end{equation}

The asymptotic form of $Z_{2r}(\om)$ as $r\rightarrow \infty$ was obtained in
\cite{BEO} and will now be used to study the phase diagram for the ASEP model.
First we outline the method by which the asymptotic form was obtained.

Notice that expanding the factor $(1- \omega \Lambda^2)^{-1}$ in \eqref{Zom} in
powers of 
$\omega$ gives an infinite series  
which is only valid for $c \le 1$ which is the point at which $\omega$ as a
function of $c$ passes through its maximum value $\frac 14$. Instead we use
\eqref{Zom} to obtain a recurrence relation. Thus noting that $$
\frac {\omega\Lambda^{2r}}{1- \omega \Lambda^2} = - \Lambda^{2r-2}+
\frac {\Lambda^{2r-2}}{1- \omega \Lambda^2}$$
and substituting in \eqref{Zom} gives
\begin{equation}
\label{rec3}
\omega Z_{2r}(\omega) = -C_{r-1} + Z_{2r-2}(\omega).
\end{equation}
where $C_{r}$ is the Catalan number, \eqref{CTCat}. Solving \eqref{rec3} with $Z_2 = (1+c)^2$ gives
\begin{equation}
Z_{2r}(\oc) = \oc^{-r}( 1+c - \sum_{j=0}^{r-1} C_j\oc^j)
\end{equation}
which on substituting for $\oc$ in terms of $c$ must give a polynomial in $c$.

For the ASEP model $\oc = \alpha(1-\alpha)$ and
\begin{align*}
\sum_{j=0}^\infty C_j \oc^j &= \frac{1-\sqrt{1-4\oc}}{2\oc} = 
\frac{1-|1-2\alpha|}{2\alpha(1-\alpha)}\\
&= \left\{
\begin{array}{ll}
\frac 1 \alpha&\alpha > \frac 12\\
\frac 1{1-\alpha}&\alpha \le \frac 12
\end{array}
\right.
\end{align*} so
\begin{align} Z_{2r}(\oc) &= \oc^{-r}\left[\frac 1 \alpha 
- (\sum_{j=0}^{\infty} C_j\oc^j-\sum_{j=r}^\infty C_j\oc^j)\right]\\
&= \oc^{-(r+1)}(1-2\alpha)\,\,\theta(1-2\alpha) + \sum_{j=r}^\infty C_j\oc^{j-r}
\end{align} where $\theta(.)$ is the unit step function. This is  \cite{BEO} 
equation (3.61).

The asymptotic form for $r\rightarrow\infty$ was obtained using $$
C_r\sim \frac {4^r}{\sqrt{\pi}r^{3/2}}$$
and approximating the sum by an integral with the result

\begin{align}\label{eq:reg2}
 Z_{2r}(\oc)\sim \left\{
\begin{array}{ll}
\frac{1-2\alpha}{\oc^{r+1}}&\alpha<\frac 12\\
\frac 2 {\sqrt{\pi}}\frac {4^r}{r^{\frac 12}}& \alpha = \frac 12\\
\frac{4^r}{\sqrt{\pi} r^{\frac 32}} \left(\log(\frac 1{4\oc})\right)^{-1}&\alpha
> \frac 12
\end{array}
\right.
\end{align} and hence
\begin{align}
\label{eq:reg2}
\oc Z_{2r+2}(\oc)\sim \left\{
\begin{array}{ll} f_<(\a)=\frac{1-2\alpha}{\oc^{r+1}}&\alpha<\frac 12\\
f_==\frac 2 {\sqrt{\pi}}\frac {4^r}{r^{\frac 12}}& \alpha = \frac 12\\
f_>(\a)=4\oc\frac{4^r}{\sqrt{\pi} r^{\frac 32}} \left(\log(\frac
1{4\oc})\right)^{-1}&\alpha > \frac 12
\end{array}
\right.
\end{align}

As $\alpha \rightarrow \frac 12$, $\log(\frac 1{4\oc})\sim \frac
{(2\alpha-1)^2}{4\alpha^2}$ and these results agree with \cite{DEHP} equations
(48)-(50).

In the phase diagram there are therefore
\begin{itemize}
\item three special regions $R_1=\{\a>\ha,\b>\ha\}$, $R_2=\{\a>\b,\b<\ha\}$,
$R_3=\{\a<\b,\a<\ha\}$

The partition function $R_3$ is obtained from that in $R_2$ by interchanging
$\a$ and $\b$.
\item three special lines $L_1=\{\a=\b<\ha\}$, $L_2=\{\b=\ha,\a>\ha\}$,
$L_3=\{\a=\ha,\b>\ha\}$ The partition function $L_3$ is obtained from that in
$L_2$ by interchanging $\a$ and $\b$.
\item a special point where the lines meet $P=\{\a=\ha,\b=\ha\}$
\end{itemize}
\begin{figure}[ht!]
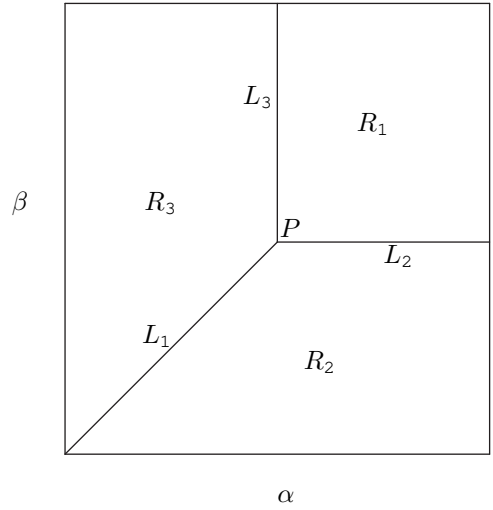

\centering
\phasediag
\caption{\it  The various regions of the ASEP phase diagram.}\label{fig:pd}
\end{figure}

Table \ref{tableO} shows the asymptotic form of $\pf_{2r}(1|1;\k_1,\k_2)$ for the above cases
\begin{table}[ht!]\label{tableO}
\begin{center}
\begin{tabular}{|c|c|c|}
\hline &&\\
$R_3:\frac{f_<(\a)}{\bar\a-\bar\b}$&$L_3:\frac{f_=}{2-\bar\b}$
&$R_1:\frac{f_>(\a)-f_>(\b)}{\bar\a -\bar\b}$\\
&&\\
\hline &&\\
&$P:4^r$&$L_2:\frac{f_=}{2-\bar \a}$\\
&&\\
\hline &&\\
$L_1:-\a^2 f_<'(\a)\sim
\frac{r(1-2\a)^2}{\oc^{r+2}}$&&$R_2:\frac{f_<(\b)}{\bar\b-\bar\a}$\\
&&\\
\hline
\end{tabular}
\end{center}
\caption{Asymptotic forms of $\pf_{2r}(1|1;\k_1,\k_2)$ in the regions defined in the phase diagram in figure \ref{fig:pd}.}
\label{tableO}
\end{table}

\section{Recurrence relations for the partition function and correlation functions.}
\subsection{Recurrence relations for the partition function.}

The various formulae for $G(z)$ when substituted in \eqref{Z11} 
yield recurrence relations for $\pf_{t}(y|1;\k_1,\k_2)$.
For example, using the identity
$$\frac 1 {(1-\oc \La^2)(1-\od \La^2)} = 1 + \frac {(\oc+\od)\La^2 - \oc \od \La^4} {(1-\oc \La^2)(1-\od \La^2)},$$
substitution in \eqref{Zomcd} and using the CT formula \eqref{CTCat} for Catalan numbers leads to,
for $r=0,1,\dots$
\begin{equation}
\label{omegarecurrence}
\pf_{2r}(1|1;\k_1,\k_2) = \frac{\oc -\od}{c-d}C_{r +1} +
(\oc+\od)\pf_{2r+2}(1|1;\k_1,\k_2) -\oc\od\,\pf_{2r+4}(1|1;\k_1,\k_2).
\end{equation}
Substitution in terms of $\k_1$ and $\k_2$ yields, for $r=2,3,\dots$
\begin{equation}
(\k_2-1)\pf_{2r}(1|1;\k_1,\k_2) = (\k_2(\k_1+\k_2)-2\k_1)\pf_{2r-2}(1|1;\k_1,\k_2)
+\k_1^2 \pf_{2r-4}(1|1;\k_1,\k_2) - \k_2 C_{r-1}
\end{equation}
which may be initialised by $\pf_0(1|1;\k_1,\k_2)=1$ and 
$\pf_2(1|1;\k_1,\k_2)=\k_1 + \k_2$.

The following identity
$$
\frac{1}{(1-\ab z/\La)(1-\bb z/\La)} = 1 + \frac{(\ab+\bb)z/\La - \ab \bb z^2/\La^2}{(1-\ab z/\La)(1-\bb z/\La)}
$$
when substituted in \eqref{Gabbb} gives, using the $CT$ formula \eqref{CTBallot} for Ballot numbers,
for $y=1,2,\dots$ and odd $t+y\ge 3$
\begin{equation}\label{abrecurrence}
\pf_{t}(y|1;\k_1,\k_2) = B_{t-1,y-2} + (\ab +\bb)\pf_{t-1}(y+1|1;\k_1,\k_2)-\ab\bb \pf_{t-2}(y+2|1;\k_1,\k_2)
\end{equation}
which relates partition functions on lines of constant $t+y$.  The partition function for $y=1$ is determined
by \eqref{omegarecurrence} and the following proposition then the determines $\pf_{t}(2|1;\k_1,\k_2)$ which
provides the initial condition for \eqref{abrecurrence}.
\begin{prop}
For $r = 1,2,\dots$
$$\ab \pf_{2r-1}(2|1;\k_1,\k_2) = \pf_{2r}(1|1;\k_1,\k_2) - \pf_{2r}(1|1;0,\bb)$$
\end{prop}
\begin{proof}
 From corollary \ref{alphabeta},
for $r = 1,2,\dots$
\begin{align*} 
\Za_{2r}(1|1;\k_1,\k_2) 
&= \sum_{m=1}^r B_{2r-m-1,m-1}\sum_{i=1}^{m}\bar \alpha^i \bar \beta^{m-i}+
\sum_{m=1}^r B_{2r-m-1,m-1}\bar \beta^{m}\\
&=\ab \sum_{m=0}^{r-1} B_{2r-m-2,m}\sum_{i=0}^{m}\bar \alpha^i \bar \beta^{m-i}+\pf_{2r}(1|1;0,\bb)
\end{align*}
and the result follows from corollary \ref{alphabeta} with $t=2r-1, y=2$.
\end{proof}

Finally, substituting the identity

\begin{equation}
\frac {1} {(1-c z^2)(1- d z^2)} = 1 +\frac
{ (c+d)z^2 - cd z^4} {(1-c z^2)(1- d z^2)}
\end{equation}
in \eqref{Gcd} and using \eqref{Z11} leads to the recurrence, for $t = 0,1,2,\dots$, and odd $t+y \ge 1$
\begin{equation}
\label{crecurrence}
\pf_{t}(y|1;\k_1,\k_2) = B_{t+1,y} + (c+d)\pf_{t}(y+2|1;\k_1,\k_2)
-cd\,\pf_{t}(y+4|1;\k_1,\k_2).
\end{equation}
which relates partition functions along lines of constant $t$ and may be initialised using the above relations
to find the partition functions for $y=1,2,3$ and $4$.

\subsection{Recurrence relations for the correlation functions of the ASEP model.}

\begin{figure}[ht!]
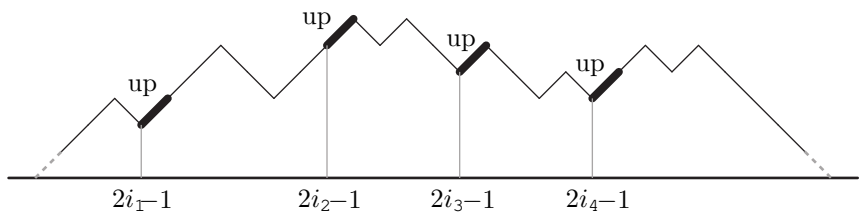

\centering
\corrpath
\caption{\it   The $n$-point correlation path interpretation ($n=4$ is shown above):
each step starting at $x=2i_k-1$, $k =1,\dots,n$ must  be an up step with weight one.}
\label{fig:corrpath}.
\end{figure}

The probability of finding particles at positions $i_1,i_2,\dots,i_n$ is, using \eqref{prob},
\begin{equation}
Pr(\tau_{i_1}=1, \tau_{i_2}=1,\dots, \tau_{i_n}=1) = \frac{G_n(i_1,i_2,\dots,i_n;r)}{\pf_{2r}(1|1;\ab\bb,\k^2)}
\end{equation}
where the un-normalised $n$-point correlation function $G_n(i_1,i_2,\dots,i_n;r)$ is given by
\begin{equation}
G_n(i_1,i_2,\dots,i_n;r) = <W|C^{i_1-1}DC^{i_2-i_1-1}D\dots 
C^{i_n-i_{n-1}-1}D C^{r-i_n}|V>.
\end{equation}
where $C=DE$.
This expression may be thought of as
replacing $C$ by $D$ in  $<W|C^r|V>$ at each of the positions $i_k, k
=1,\dots,n$ which is equivalent to 
replacing the $E_j$ matrix  in a $C_j=D_jE_j$ product by a unit matrix. Thus in the path
representations $G_n(i_1,i_2,\dots,i_n;r)$ is obtained by modifying the allowed step definition
such that for $k =1,\dots,n$, the step $s_k \equiv e_{2i_k}$ (beginning at $x=2i_k-1$ and ending at $x=2i_k$) 
is always an up step and has  weight $1$. We will say that $s_k$ is a {\it forced up step}.
This is illustrated in figure \ref{fig:corrpath}.

\subsubsection{The case $\alpha=\beta=1$}

In the case $\alpha=\beta=1$, or $\k_1=\k_2=1$, it is shown in
\cite{DEHP} that
\begin{equation}\label{Zval}
\pf_{2r}(1|1;1,1)\equiv\,<W|C^r|V>|_{\a=\b=1} = C_{r+1},
\end{equation}
a Catalan number. This may also be seen in terms of the third path representation (corresponding to the third matrix representation of
section \ref{matrepsec}). $<W_3|(D_3 E_3)^r|V_3>|_{\a=\b=1}$ is just the  total number of
One Up paths of length $2r$ and these biject to Dyck paths of length $2r+2$ by adding an up step at the 
beginning and a down step at the end of each path (the dotted steps in figure \ref{fig:corrpath}).
The result follows since number of Dyck paths of length 
$2p$ is well known to be the Catalan number $C_{p}$. The result also follows
from \eqref{Zac}  by setting $c=d=0$ and noting that $B_{2r+1,1} = C_{r+1}$
which is the $m=0$ term.

\begin{figure}[ht!]
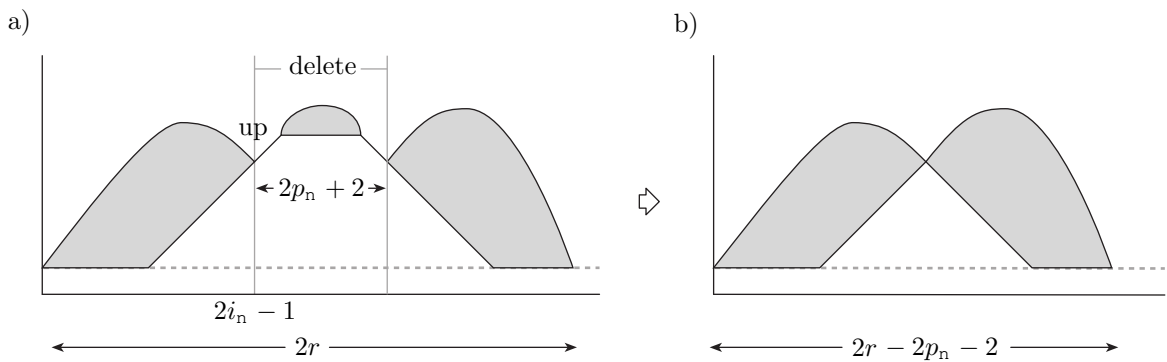

\centering
\correlation
\caption{\it  a)  Schematic path picture for the correlation functions for $\alpha=\beta=1$, showing the bubble following the last forced up step. b) Removing the bubble  leaves a shorter Dyck path }
\label{bubble1}
\end{figure}

It is also shown in \cite{DEHP}, equation (88) that 
\begin{equation}
\label{recur} G_n(i_1,i_2,\dots,i_n;r) = 
\sum_{p_n=0}^{r-i_n}C_{p_n}G_{n-1}(i_1,i_2,\dots,i_{n-1};r-p_n-1)
\end{equation}
This may be derived combinatorially as follows. Again we use the third path representation and to avoid special 
cases we imagine that the paths are extended to $y=0$
by a further down step. For each
path which  contributes to $G_n(i_1,i_2,\dots,i_n;r)$ we determine a subpath $\om_n$ which starts
with the last forced up step, $s_n$, and ends when the path returns to the same height
for the first time. This subpath contains a Bubble which we suppose has length $2p_n$ so that the
subpath has length $2p_n+2$-- see figure \ref{bubble1}a.  Immediate 
return corresponds to $p_n=0$ and the maximum value of $p_n$ is determined by the condition,
$(2i_n-1)+(2p_n+2)= 2r+1$, that there are no further steps beyond $\om_n$.   

We can now define a new path obtained by deleting $\om_n$ and joining the two (possibly empty)
resulting sub-paths which remain. This path contributes to
$G_{n-1}(i_1,i_2,\dots,i_{n-1};r-p_n-1)$. The result  follows by partitioning
the paths contributing to  $G_n$  according to the value of $p_n$. For a given
value of $p_n$ the number of configurations is therefore the product $G_{n-1}$ 
and the number of configurations
of $\om_n$ which is equal to the number, $C_{p_n}$, of  Dyck paths of length $2p_n$. 
 
For $k=1,2,\dots n$, let
\begin{equation}
q_k= n - k + \sum_{j=k}^np_j
\end{equation}
then as pointed out in \cite{DEHP}, equation \eqref{recur}  may be iterated or,
combinatorially, $n$ Bubbles may be removed, to give the explicit formula
\begin{equation}
G_n(i_1,i_2,\dots,i_n;r)= \sum_{p_1\geq 
0}\dots\sum_{p_n\geq 0} C_{p_1}C_{p_2}\dots C_{p_n}C_{r-q_1}
\end{equation}
where the upper limits are $p_k = r-i_k - q_{k+1}$ with $q_{n+1}=0$.

This formula was previously conjectured by Derrida and Evans \cite{DE} on the basis of computer
calculations up to $r=10$.

\subsubsection{General $\alpha$ and $\beta$.}

Equation (45) of \cite{DEHP} is the case $n=2$ of the following proposition
which we now prove using a lattice path representation. An algebraic proof was
given in \cite{DEHP}.
\begin{prop} For $1\le i_1 < i_2 < \dots < i_n \le r-1$
\begin{align}\label{eq:bcor} G_n(i_1,i_2,\dots,i_n;r) = \sum_{p=0}^{r-i_n-1}
&C_p G_{n-1}(i_1,i_2,\dots,i_{n-1};r-p-1)\notag\\
&+G_{n-1}(i_1,i_2,\dots,i_{n-1};i_n-1) \sum_{p=2}^{r-i_n+1}
B_{2r-2i_n-p,\,p-2}\bb^p.
\end{align}
\end{prop} {\it Notes:} 
\begin{itemize}
\item When $\a = \b =1$ this reduces to \eqref{recur} since using $$
\sum_{p=2}^{k+1} B_{2k-p,p-2} = C_k $$
the second sum of \eqref{eq:bcor} is just the missing term $p=r-i_n$ of the
first sum. 
\item
 In constructing a proof it was found that the first representation in terms of
Jump-Paths was simpler to use than the third which we used in the special case
of the previous section. It was explained after definition \ref{def:js} , that Jump-Step paths (representing 
$Z_{2r}$) never intersect $y=0$. However this is not the case in calculating  $G_n(i_1,i_2,\dots,i_n;r)$
since  the paths  are modified by the forced
up steps. This allows $y=0$ to be visited and then a forced up step returns the path to
$y=1$. Notice that the down step leading to $y=0$ has a weight
$\bb$. 
 
\end{itemize}

\begin{figure}[ht!]
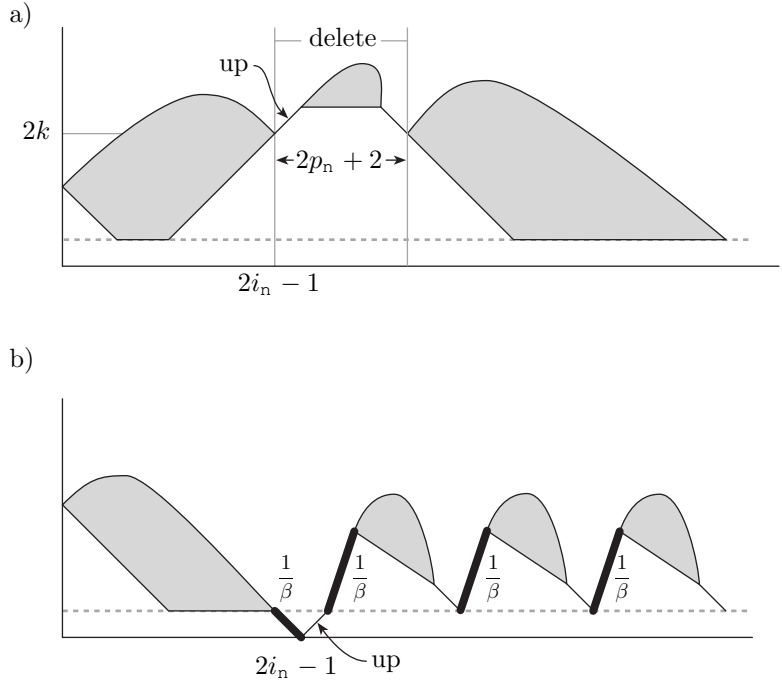

\centering
\gencor
\caption{\it  a) Case A: Last forced up step starts at  $y = 2k, k \ge 1$.  b) Case B: 
Last forced up step starts at $y=0$. Note, the ``bubbles'' represent Jump Step subpaths. }

\label{bubble2}
\end{figure}

{\it Proof:} \quad Partition the modified Jump-Step configurations according as the last
forced up step, $s_n$, which starts at height $y = 2k, k \ge 1$ (case A) or $y=0$
(case B)  -- see figure \ref{bubble2}. 

\paragraph{Case A:} A subpath $\om_n$ may be identified in exactly the same way
as in the case $\a=\b=1$ but in this case the Bubble of length $2p$ obtained by deleting the first
and last steps of $\om_n$ is not an elevated Dyck path since it may contain jump steps.
However the number of configurations of $\om_n$ is still equal to $C_p$ (see below) for any value of $k$
and since $\om_n$ avoids $y=1$ its steps are unweighted.
Thus on partitioning the paths according to the length
$(2p+2, p = 0,1,\dots r-i_n-1)$ of $\om_n$ a factor $C_p$ may be removed from the sum over paths 
having the same value of $p$.  
When $\om_n$ is deleted the remaining steps form a weighted path of length
$2r-2p-2$ which has only $n-1$ forced up steps. Summing over configurations of this path
for given $k$ and then summing over $k\ge1$ gives $G_{n-1}(i_1,i_2,\dots,i_{n-1};r-p-1)$.
The first term of the proposition formula is thus derived provided that we can show that the number
of configurations of $\om_n$ is $C_p$. 

Now the paths $\om_n$ of length $2p+2$ biject to $\Jump_{2p;1}$ by vertical translation through distance $2k$
and from \eqref{jweights} and \eqref{w4}
\begin{equation}
|\Jump_{2p;1}| = Z_{2p}^{(1)}|_{\ab=0,\bb=1} = Z_{2p}^{(3)}|_{\k_1=0, \k_2= 1} = C_p.
\end{equation}
The last equality follows since when $\k_1=0$ the One Up paths which visit $y=0$ have zero weight and
the remaining paths, which have weight $1$ when
$\k_2=1$, biject to Dyck paths by vertical translation through unit distance. This result is also proved combinatorially in \cite{BER}.

\paragraph{Case B:} The weighted sum over paths may be factorised into three
parts. 
\begin{itemize}
\item[(i)] A factor which arises from the subpath consisting of the first $2i_n-2$ steps. The subpath
ends at $y=1$ and has only $n-1$ forced up steps, therefore the sum over these
subpaths yields 
$G_{n-1}(i_1,i_2,\dots,i_{n-1};i_n-1)$. 
\item[(ii)] A factor $\bb$ which arises from the next two steps which visit
$y=0$ and return to $y=1$. The second of these is the last forced up step $s_n$ having weight $1$.
\item[(iii)] A factor arising from the subpath consisting of the remaining
$2r-2i_n$ steps which is a jump step path beginning and ending at $y=1$, avoiding $y=0$.  
The weighted sum over these paths is obtained by setting $\ab=0$ in the normalising factor
for paths of length $2r-2i_n$, thus using
corollary \ref{alphabeta}
$$Z_{2r-2i_n}|_{\ab =0} = \sum_{m=1}^{r-i} B_{2r-2i_n-m-1, \,m-1}\bb^m.$$

\end{itemize}

The product of these three factors yields the second term of the formula.

\section{Conclusion}

 We have shown that the normalisation of the ASEP can be interpreted as various lattice path problems. The lattice path problems can then be solved using  the constant term method (CTM).  The   combinatorial nature of the CTM enables us to interpret the coefficients of the normalisation polynomials as various un-weighted lattice path problems -- usually as Ballot paths.  One particular form has a natural interpretation as an equilibrium polymer chain adsorption model. The ``$\omega$'' form of the normalisation is particularly suited to finding the asymptotic expansion of normalisation and hence the phase diagram. We also formulate a combinatorial interpretation of the correlation functions.

The lattice path interpretations enable  us to make connections with many other models. In particular, because of the strong combinatorial nature of the  CTM  we are able to find a  new ``canonical''  lattice path representation. In a further paper \cite{BDR} we show that this representation leads to an understanding of a non-equilibrium model (the ASEP model) in terms of a related equilibrium 
polymer model. Also, having extended the polymer chain model so that the endpoints have arbitrary displacements from the surface
the method of Gessel and Viennot \cite{GV1}, \cite{GV2} (see also \cite{KM}) may be used 
to express the partition function of a network of non-intersecting paths as a
determinant. In particular,  the case of two paths gives the partition function for a vesicle model  with a two parameter interaction with a surface. A bijection between these vesicles and compact percolation clusters \cite{DK} then enables an analysis of the properties of the clusters attached to damp wall to be made. These applications will also be the subject of a subsequent publication.

It is also of combinatorial interest to understand how the various path problems might be related. Clearly they are related algebraically as they are all just representations of the same algebra (and thus related by different similarity transformations).   In a subsequent paper, \cite{BER}, we show combinatorially (using bijections and involutions) 
how the various path representations are combinatorially  equivalent.

\section{Acknowledgements}
Financial support from the Australian Research Council is
gratefully acknowledged.  JWE is also grateful for the kind
hospitality provided by the University of Melbourne and RB
is also grateful for the kind hospitality provided by Royal
Holloway College, London University. We would like to thank Andrew Rechnitzer and Aleks Owczarek for reading the manuscript and providing useful comments. 

\newpage
{\bf \Large APPENDIX}
\appendix
\section{Specimen partition functions.}
\begin{align}
\pf_0(1|1;\k_1,\k_2) =&1\\
\pf_2(1|1;\k_1,\k_2) =&{{\kappa }_1} + {{\kappa }_2}\\
\pf_4(1|1;\k_1,\k_2) =&{{{\kappa }_1}}^2 + {{\kappa }_2} + 2\,{{\kappa }_1}\,{{\kappa }_2} + {{{\kappa }_2}}^2\\
\pf_6(1|1;\k_1,\k_2) =&{{{\kappa }_1}}^3 + 2\,{{\kappa }_2} + 2\,{{\kappa }_1}\,{{\kappa }_2} + 3\,{{{\kappa }_1}}^2\,{{\kappa }_2} + 
  2\,{{{\kappa }_2}}^2 + 3\,{{\kappa }_1}\,{{{\kappa }_2}}^2 + {{{\kappa }_2}}^3\\
\pf_8(1|1;\k_1,\k_2)= &{{{\kappa }_1}}^4 + 5\,{{\kappa }_2} + 4\,{{\kappa }_1}\,{{\kappa }_2} + 3\,{{{\kappa }_1}}^2\,{{\kappa }_2}\notag\\ 
&+ 
  4\,{{{\kappa }_1}}^3\,{{\kappa }_2} + 5\,{{{\kappa }_2}}^2 + 6\,{{\kappa }_1}\,{{{\kappa }_2}}^2 + 
  6\,{{{\kappa }_1}}^2\,{{{\kappa }_2}}^2 + 3\,{{{\kappa }_2}}^3 + 4\,{{\kappa }_1}\,{{{\kappa }_2}}^3 + 
  {{{\kappa }_2}}^4 \\
&\notag\\
\pf_0(3|1;\k_1,\k_2)=&0\\
 \pf_2(3|1;\k_1,\k_2) =&1\\
 \pf_4(3|1;\k_1,\k_2) =&2+\k_1 + \k_2 \\
\pf_6(3|1;\k_1,\k_2) =&5 + 2\,{{\kappa }_1} + {{{\kappa }_1}}^2 + 3\,{{\kappa }_2} + 2\,{{\kappa }_1}\,{{\kappa }_2} + {{{\kappa }_2}}^2  \\
\pf_8(3|1;\k_1,\k_2) = &14 + 5\,{{\kappa }_1} + 2\,{{{\kappa }_1}}^2 + {{{\kappa }_1}}^3 + 9\,{{\kappa }_2} + 
  6\,{{\kappa }_1}\,{{\kappa }_2} + 3\,{{{\kappa }_1}}^2\,{{\kappa }_2} + 4\,{{{\kappa }_2}}^2 \notag\\
  &+ 3\,{{\kappa }_1}\,{{{\kappa }_2}}^2 + {{{\kappa }_2}}^3\\
&\notag\\
 \pf_t(1|3;\k_1,\k_2)=&\k_2\pf_t(3|1;\k_1,\k_2)\\
&\notag\\
\pf_0(3|3;\k_1,\k_2) =&1\\
 \pf_2(3|3;\k_1,\k_2)=&2\\
 \pf_4(3|3;\k_1,\k_2) =&5+\k_2\\
 \pf_6(3|3;\k_1,\k_2) =&14 + 4\,{{\kappa }_2} + {{\kappa }_1}\,{{\kappa }_2} + {{{\kappa }_2}}^2\\
\pf_8(3|3;\k_1,\k_2) =&42 + 14\,{{\kappa }_2} + 4\,{{\kappa }_1}\,{{\kappa }_2} + {{{\kappa }_1}}^2\,{{\kappa }_2} + 5\,{{{\kappa }_2}}^2 + 
  2\,{{\kappa }_1}\,{{{\kappa }_2}}^2 + {{{\kappa }_2}}^3
  \end{align}

\section{Constructive proof of proposition \ref{prop1}.}

The formula 
$$\pf_t(y|y^i;\k_1,\k_2)=CT[\La^tz^y (\bar z^{y^i} + U(z) z^{y^i})]$$
clearly satisfies the general equation \eqref{general} and the initial condition \eqref{Zdef2} 
provided that 
\begin{equation}\label{condition}
CT[z^{y+y^i}U(z)]=0.
\end{equation}
In order to satisfy the boundary condition \eqref{modZ}

$$CT[\La^t z(\bar z^{y^i} + U(z) z^{y^i})] = \ka_1CT[\La^{t-2}z(\bar z^{y^i} + U(z) z^{y^i})]+
\ka_2 CT[\La^{t-1}z^2 (z^{y^i} + U(z) z^{y^i})]$$
or
$$CT[\La^{t-2}\bar z^{y^i-1}(\La^2 -\ka_1-\ka_2 z \La)]= -CT[\La^{t-2}V(z) z^{y^i-1}]$$
where
$$V(z) = (1-(\bar \ka_1+\bar \ka_2)z^2 -\bar \ka_2 z^4) U(z).$$
Replacing $z$ by $\bar z$ everywhere under the CT operation leaves the value unchanged so 
$$V(z) = - (\La^2-\ka_1-\La \bar z \ka_2) = \bar z^2(\ka_2-1 + (\bar\ka_1+\bar\ka_2)z^2 -z^4)$$
and \eqref{condition} is satisfied provided that $y+y^i >2$ or $y^i>1$ since $y\ge 1$. Hence in this case
$y^i>1$
$$U(z) = \bar z^2 (-1 + \ka_2 G(z)).$$

When $y^i=1$ the boundary condition is satisfied if
$$CT[\La^{t-2}(\La^2 -\ka_1-\ka_2 z \La)]
\equiv CT[\La^{t-2}(\bar z^2-2+z^2 -\ka_1-\ka_2 (1+z^2))]= -CT[\La^{t-2}V(z)]$$
and replacing $\bar z^2$ by $z^2$ in the second expression gives
$$V(z) = \bar\ka_1+\bar\ka_2 +\bar \ka_2 z^2-z^2$$
which satisfies \eqref{condition} and hence
$$U(z) =\bar z^2(-1 +G(z)).$$


\end{document}